\documentclass[journal,twoside]{IEEEtran}
\usepackage{cite}
\usepackage{graphicx}
\usepackage{epstopdf}
\usepackage[cmex10]{amsmath}\interdisplaylinepenalty=2500
\usepackage[caption=false]{subfig}
\usepackage{amssymb,bm,upgreek}

\newtheorem{lemma}{Lemma}
\newtheorem{proposition}{Proposition}
\newtheorem{remark}{Remark}

\begin{document}

\title{On the Pilot Contamination Attack in Multi-Cell Multiuser Massive MIMO Networks}

\author{Noman~Akbar,~\IEEEmembership{Member,~IEEE,}
        Shihao~Yan,~\IEEEmembership{Member,~IEEE,}\\
        Asad~Masood~Khattak,~\IEEEmembership{Member,~IEEE,}
        and Nan~Yang,~\IEEEmembership{Senior Member,~IEEE}

\thanks{N. Akbar and N. Yang are with the Research School of Electrical, Energy and Materials Engineering, Australian National University, Canberra, ACT 2601, Australia (e-mails: \{noman.akbar, nan.yang\}@anu.edu.au).}
\thanks{S.~Yan is with the School of Engineering, Macquarie University, Sydney, NSW 2109,  Australia (e-mail: shihao.yan@mq.edu.au).}
\thanks{A.~M.~Khattak is with the College for Technological Innovation, Zayed University, Abu Dhabi, UAE (e-mail: asad.khattak@zu.ac.ae).}
\thanks{This work was presented in part~\cite{Akbar2017a} at the IEEE GlobeCOM Workshop on Trusted Communications with Physical Layer Security, Singapore, Dec. 2017.}
\thanks{The corresponding author is Shihao Yan.}}

\markboth{ACCEPTED BY IEEE Transactions on Communications}{Akbar \MakeLowercase{\textit{et al.}}:On the Pilot Contamination Attack in Multi-Cell Multiuser Massive MIMO Networks}

\maketitle

\begin{abstract}
In this paper, we analyze pilot contamination (PC) attacks on a multi-cell massive multiple-input multiple-output (MIMO) network with correlated pilots. We obtain correlated pilots using a user capacity-achieving pilot sequence design. This design relies on an algorithm which designs correlated pilot sequences based on signal-to-interference-plus-noise ratio (SINR) requirements for all the legitimate users. The pilot design is capable of achieving the SINR requirements for all users even in the presence of PC. However, this design has some intrinsic limitations and vulnerabilities, such as a known pilot sequence and the non-zero cross-correlation among different pilot sequences. We reveal that such vulnerabilities may be exploited by an active attacker to increase PC in the network. Motivated by this, we analyze the correlated pilot design for vulnerabilities that can be exploited by an active attacker. Based on this analysis, we develop an effective active attack strategy in the massive MIMO network with correlated pilot sequences. Our examinations reveal that the user capacity region of the network is significantly reduced in the presence of the active attack. Importantly, the SINR requirements for the worst-affected users may not be satisfied even with an infinite number of antennas at the base station.
\end{abstract}


\section{Introduction}

\IEEEPARstart{M}{assive} multiple-input multiple-output (MIMO) has been widely acknowledged as an essential enabler for the next-generation wireless networks. In massive MIMO networks, base stations (BSs) are equipped with a huge number of antennas to offer numerous benefits over regular MIMO, such as simpler power control \cite{Marzetta2016}, higher spectral efficiency, and higher energy efficiency \cite{Bjornson2014,Akbar2018}. A key benefit of massive MIMO is that the wireless channels become increasingly orthogonal when the number of antennas at BS increases \cite{Narasimhan2014}. Thus, recent experimental results have suggested that massive MIMO offers a major improvement in performance towards achieving a thousand time increased data rate as compared to the 4G networks, which confirms the important role of massive MIMO in the 5G era \cite{Malkowsky2017,Zhang2018,Martinez2017}.

Pilot contamination (PC) is one of the key performance limiting factors for unlocking the full potential offered by massive MIMO \cite{Elijah2016,Jose2011,Akbar2016,Akbar2016b,Shen2015}. PC stems from the reuse of pilot sequences in the massive MIMO network. Consequently, the pilot sequences assigned to different users in the network are non-orthogonal \cite{Akbar2016,Akbar2016b}. We highlight that the next-generation wireless networks are required to support a large number of high-mobility users. As such, it is not possible to assign orthogonal pilot sequences to all the users in the network. Also, the short duration of the channel coherence interval restricts the use of a large number of orthogonal pilot sequences in the massive MIMO network. Consequently, there are limited orthogonal pilot sequences available in massive MIMO networks.

Considering the fact that PC degrades the performance of the massive MIMO network, a number of solutions have been proposed to compensate for this degradation \cite{Elijah2016,Lu2014}. Such solutions can be broadly grouped into five categories: 1) the protocol-based method which restricts the simultaneous transmission from the users having the same pilot sequence or wisely assigns pilot sequences among users to alleviate PC \cite{Fernandes2013,Zhu2015B,Ahmadi2015}; 2) the precoding-based method which relies on specifically designed precoders to reduce the interference caused by PC \cite{Jose2011,Ashikhmin2018}; 3) the angle-of-arrival (AoA)-based method which mitigates the interference from the users having the same pilot sequence and mutually non-overlapping AoAs \cite{Akbar2018a,Yin2013}; 4) the blind method which partitions the signal space into desired signal subspace and interference signal subspace and then develops algorithms to reduce the interference from the latter \cite{Muller2014,Hu2016}; and 5) the pilot sequence design methods which aim at designing pilot sequences such that the PC does not severely impacts the network performance \cite{Akbar2016,Akbar2016b,Shen2015}. It is worth mentioning that most conventional methods assumed orthogonal pilot sequences to perform PC analysis \cite{Akbar2018a,Fernandes2013,Jose2011,Yin2013}. However, this assumption may not be practical in realistic massive MIMO networks. Fortunately, pilot sequence design methods relax the assumption of strict orthogonality between the pilot sequence set \cite{Akbar2016,Akbar2016b,Shen2015}. As such, the pilot sequences set is non-orthogonal and every pilot sequence has a non-zero cross-correlation with other pilot sequences. Accordingly, the pilot sequence design methods aim at minimizing the cross-correlation such that the network achieves its performance target, e.g., signal-to-interference-plus-noise ratio (SINR) requirements. Although pilot sequence design methods provide an adequate solution to the PC problem, the use of correlated pilots makes the massive MIMO network more susceptible to PC attacks by an adversary \cite{Akbar2017a}.

We highlight that PC attack may be difficult to detect in a massive MIMO network \cite{Kapetanovic2015,Zhou2012}. Recently, several methods have been proposed for PC attack detection \cite{Tugnait2015,Tugnait2018,Wang2018}. PC attacks can be detected using self-contamination \cite{Tugnait2015,Tugnait2018} or via random channel training \cite{Wang2018}. If an attacker avoids detection and countermeasures, it may successfully modify the downlink precorder at BSs for legitimate users with PC attack \cite{Zhou2012}. Different from \cite{Zhou2012}, we consider a network with multiple Bobs and Eves. Furthermore, we analyze the optimal pilot selection at Eves for PC attack when the pilot sequences used in the network are correlated. As such, the results, analysis, and discussions in this manuscript cannot be obtained directly from \cite{Zhou2012}. Notably, it is possible to increase the adversity of PC attack with multiple Eves \cite{Huang2018}. As such, PC attack has large potential to severely degrade the performance of massive MIMO networks.

In this paper, we analyze the user capacity-achieving pilot sequence design from the perspective of an active attacker. We highlight that the user capacity-achieving pilot sequence design method composes correlated pilots for all the users in the network \cite{Akbar2016,Akbar2016b,Shen2015}. The user capacity-achieving pilot sequence design determines the user capacity region of the network under PC, where the user capacity-region signifies the range for the SINR requirements that can be satisfied by using the user capacity-achieving pilot sequence design. This design also recommends a downlink power allocation method to minimize the interference between different users in the network. The major advantage of using the user capacity-achieving pilot sequence design is that it can satisfy a diverse range of SINR requirement for all the users in the network as long as the SINR requirements lie inside the user capacity region of the massive MIMO network. In this paper, we demonstrate the vulnerabilities in the user capacity-achieving pilot sequence design. As such, an active attacker with limited knowledge about the network configuration and parameters can successfully reduce the achievable SINR of the users in the network such that they may no longer be satisfied with the user capacity-achieving pilot sequence design. Furthermore, we propose an active attacking strategy for increasing PC in the network. Additionally, we demonstrate the limitations of using correlated pilots and reveal that the use of correlated pilots makes a network prone to active attacks. To this end, we analyze the network performance under the active PC attack.

The major contributions and novelty of this work are summarized as follows:
\begin{itemize}
\item
We analyze and identify the limitations of using correlated pilots in a multiuser massive MIMO network where the correlated pilots are designed using the user capacity-achieving pilot sequence design \cite{Akbar2016,Akbar2016b,Shen2015}. To this end, we first derive analytical expressions to demonstrate the reduction in the user capacity region in a massive MIMO network under PC attack. We then demonstrate that the SINR requirements for all the users in the network can no longer be satisfied with the user capacity-achieving pilot sequence design in a network under PC attack.
\item
We propose a PC attack strategy on the user capacity-achieving pilot sequence design in the considered multiuser massive MIMO network. We show that the user capacity region achieved by the user capacity-achieving pilot sequence design is significantly reduced by the PC attack. Thus, a diverse range of SINR requirements for users are no longer supported. Importantly, the SINR requirements for some users cannot be guaranteed even with an infinite number of antennas at the BS in the network under PC attack.
\item
We analyze the structure of pilot sequences designed by the user capacity-achieving pilot sequence design. We demonstrate that the structure of the pilot sequences provides useful information to an active attacker, which can then be exploited for a PC attack.
\end{itemize}

We present numerical results to demonstrate the impact of PC attack under varying network parameters. Our numerical results demonstrate that PC attack reduces the SINR requirements for some users such that they can no longer be satisfied even with an unlimited number of antennas at the BS. Furthermore, the PC attack reduces the SINR for other users such that additional antennas are required at the BS to fulfill their SINR requirements during the downlink transmission phase.

\emph{Notations}: Vectors and matrices are denoted by lower-case and upper-case boldface symbols, respectively. $(\cdot)^{T}$ denotes the transpose, $(\cdot)^{H}$ denotes the Hermitian transpose, $\otimes$ denotes the Kronecker product, $\mathbb{E}[\cdot]$ denotes the mathematical expectation, $\|\cdot\|$ denotes the ${l_{2}}$ norm, $\textrm{tr}(\cdot)$ denotes the trace operation, and $\text{var}(\cdot)$ denotes the variance operation.

\section{System Model}\label{Sec:System_Model}
\begin{figure}[!t]
\centering
\includegraphics[width=0.8\columnwidth]{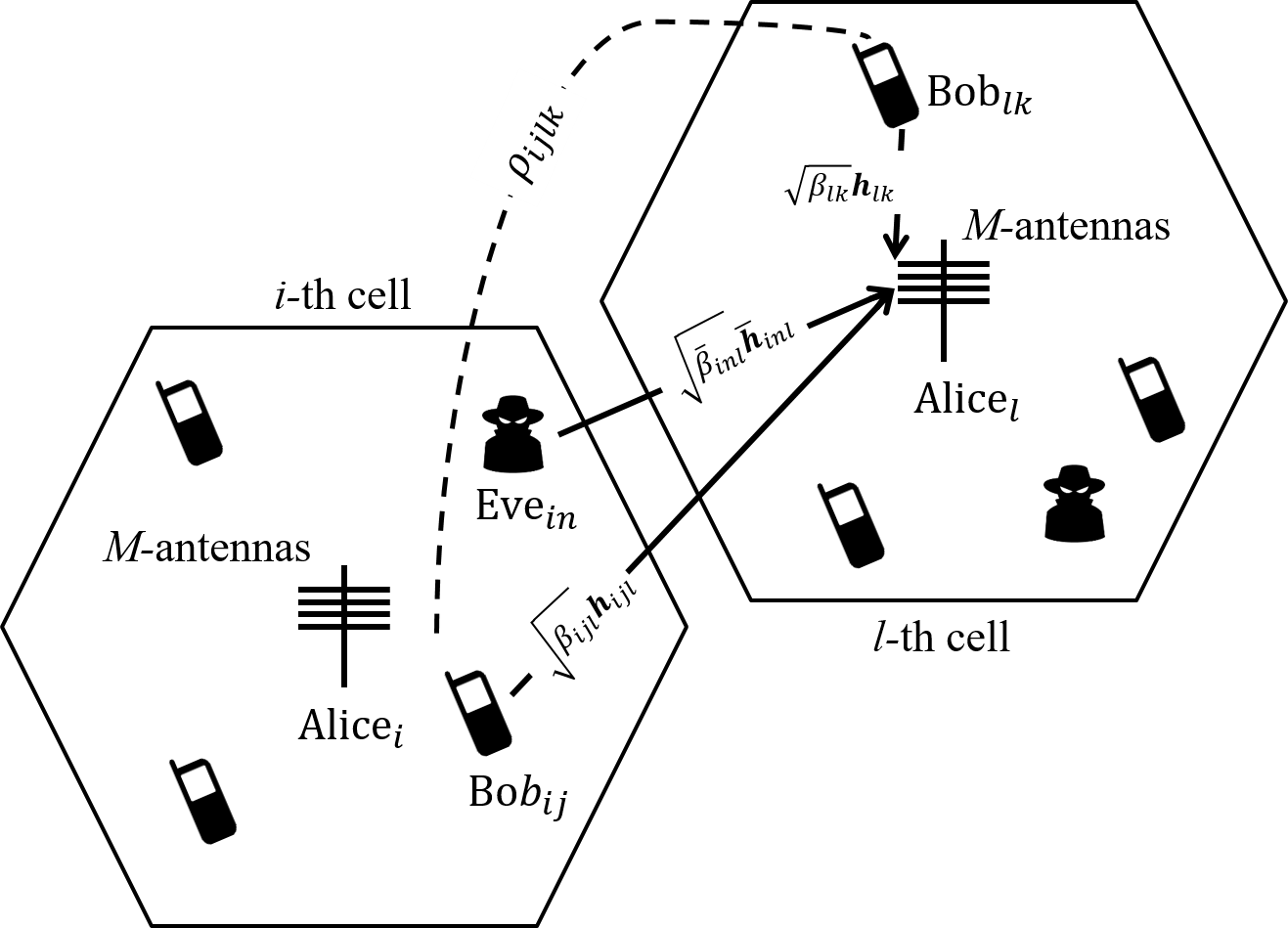}
\caption{Illustration of a multi-cell massive MIMO network with active attackers.}
\label{achievable SINR}
\end{figure}

We consider a multi-cell massive MIMO network consisting of $L$ cells with $N$ single-antenna active attackers in each cell, denoted by Eves. Each cell consists of an $M$-antenna array BS, denoted by Alice, and $K$ legitimate single-antenna users, denoted by Bobs. We assume that the channels between various nodes (i.e., Alice, Bobs, and Eves) encounter both small-scale and large-scale propagation effects. We express the channel between the $j$-th legitimate user in the $i$-th cell, $\textrm{Bob}_{ij}$ and Alice in the $l$-th cell, $\textrm{Alice}_l$, as $\mathbf{g}_{ijl}=\sqrt{\beta_{ijl}}\mathbf{h}_{ijl}$, where $i\in\{1,\cdots,L\}$, $j\in\{1,\cdots,K\}$, and $l\in\{1,\cdots,L\}$. Here, we denote the small-scale propagation vector between $\textrm{Alice}_l$ and $\textrm{Bob}_{ij}$ by $\mathbf{h}_{ijl}=\left[{h}_{ijl1},\dotsc,{h}_{ijlm},\dotsc,{h}_{ijlM}\right]$, the elements of which are Rayleigh distributed small-scale propagation coefficients with zero mean and unit variance, i.e., ${h}_{ijlm}\sim\mathcal{CN}\left(0,1\right)$. We denote the distance-dependent large-scale propagation coefficient between $\textrm{Bob}_{ij}$ and $\textrm{Alic}_l$ by $\beta_{ijl}$. Similarly, the channel between $\textrm{Alice}_l$ and the $n$-th Eve in the $i$-th cell, $\textrm{Eve}_{in}$, is expressed as $\bar{\mathbf{g}}_{inl}=\sqrt{\bar{\beta}_{inl}}\bar{\mathbf{h}}_{inl}$, where $n\in\{1,\cdots,N\}$ and $\bar{\beta}_{inl}$ and $\bar{\mathbf{h}}_{inl}$ denote the distance-dependent large-scale propagation coefficient and small-scale propagation vector between $\textrm{Alice}_l$ and $\textrm{Eve}_{in}$, respectively. Additionally, we assume that the number of Bobs in the network is greater than the length of the pilot sequences, i.e., $K>\tau$. As such, both the intra-cell and the inter-cell PC exists in the network \cite{Shen2015,Chen2016,Mai2018}.

The communication between Bobs and Alice consists of two phases: (i) the uplink channel estimation from Bobs to Alice and (ii) the downlink transmission from Alice to Bobs. In the uplink channel estimation phase, Bobs send their pre-assigned pilot sequences to the same-cell Alice for channel estimation. We assume that the network operates in the time division duplex (TDD) mode, indicating that the uplink and the downlink channels are assumed to be reciprocal. As such, the uplink channels and the downlink channels remain the same during a channel coherence interval. Based on the assumption of channel reciprocity, Alice utilizes the uplink channel estimates in the downlink transmission phase \cite{Zhang2015}.

\subsection{Uplink Channel Estimation Under Pilot Contamination Attack}

We now examine the impact of PC attack on the uplink channel estimation phase. In this phase, $\textrm{Alice}_l$ obtains the uplink channel estimates with the aid of the pilot sequences transmitted from Bobs in the same cell. At the beginning of each channel coherence interval, Bobs send their pre-assigned pilot sequences to $\textrm{Alice}_l$. We consider that $\textrm{Eve}_{in}$ has some knowledge of the pilot sequences used in the network. As such, $\textrm{Eve}_{in}$ transmits a pilot sequence $\bar{\mathbf{s}}_{in}$ during the uplink channel estimation phase such that the correlation between $\bar{\mathbf{s}}_{in}$ and $\mathbf{s}_{ij}$ is non-zero, where $\mathbf{s}_{ij}$ is the pilot sequence assigned to $\textrm{Bob}_{ij}$.\footnote{We highlight that in the absence of the prior knowledge about pilot sequences, $\textrm{Eve}_{in}$ can transmit a random pilot sequence, which is considered as a jamming attack in the massive MIMO network \cite{Do2017,Pirzadeh16}.} The uplink pilot transmission vector received at $\textrm{Alice}_l$ in the presence of $N$ Eves is represented as
\begin{align}\label{rec_pilot_eve}
\bar{\mathbf{y}_l}\!=\! \sum_{i=1}^{L}\sum_{j=1}^{K}\sqrt{p_{ij}\beta_{ijl}}{\mathbf{S}_{ij}}\mathbf{h}_{ijl} \!+\! \sum_{i=1}^{L}\sum_{n=1}^{N}\sqrt{\bar{p}_{in}\bar{\beta}_{inl}}{\bar{\mathbf{S}}_{in}}\bar{\mathbf{h}}_{inl} \!+\! \mathbf{z},
\end{align}
where $p_{ij}$ denotes the pilot transmit power at $\textrm{Bob}_{ij}$, $\mathbf{S}_{ij}={\mathbf{s}_{ij}}\otimes\mathbf{I}_{M}$ denotes the pilot sequence matrix for $\textrm{Bob}_{ij}$, $\bar{p}_{in}$ is the transmit power at $\textrm{Eve}_{in}$, $\bar{\mathbf{S}}_{in}={\bar{\mathbf{s}}_{in}}\otimes\mathbf{I}_{M}$ is the pilot sequence matrix for $\textrm{Eve}_{in}$, and $\mathbf{z}$ denotes the additive white Gaussian noise (AWGN) at $\textrm{Alice}_l$ with zero mean and variance of $\sigma_z$, i.e., each element of $\mathbf{z}$ follows $\mathcal{CN}(0,\sigma_z)$. We assume that the pilot sequences assigned to different Bobs in the network are correlated. Specifically, the correlation between pilot sequences assigned to $\textrm{Bob}_{ij}$ and $\textrm{Bob}_{lk}$ is given by $\rho_{ijlk}=\mathbf{s}_{lk}^{T}\mathbf{s}_{ij}$.
Afterwards, $\textrm{Alice}_l$ obtains the least-square (LS) channel estimate of the channel from $\textrm{Bob}_{lk}$ to $\textrm{Alice}_l$ by multiplying \eqref{rec_pilot_eve} with the pilot sequence matrix $\mathbf{S}_{ij}$ as \cite{Akbar2016,Shen2015,Akbar2018}.
\begin{align}\label{channel_estimate_eve_post}
\hat{\bar{\mathbf{h}}}_{lkl}&=\sqrt{p_{lk}\beta_{lkl}}\mathbf{h}_{lkl}
+\sum_{i=1 \atop (i,j) \neq}^L \sum_{j=1 \atop (l,k)}^{K}\sqrt{p_{ij}\beta_{ijl}}\rho_{ijlk}\mathbf{h}_{ijl}\notag\\
&+\sum_{i=1}^{L}\sum_{n=1}^N\sqrt{\bar{p}_{in}\bar{\beta}_{inl}}\rho_{inlk}\bar{\mathbf{h}}_{inl}+\mathbf{z},
\end{align}
where $\rho_{inlk}=\mathbf{s}_{lk}^{T}\bar{\mathbf{s}}_{in}$. We note that when $K>\tau$, we have $\rho_{ijlk}\neq0$ \footnote{We highlight that utilizing uncorrelated pilot sequences during the uplink transmission phase, i.e., $\tau=K$, is a specific case of \eqref{channel_estimate_eve_post}, where $\rho_{ijlk} = 1$ for $j=k$ and $\rho_{ijlk} = 0$ for $j\neq k$.}. Consequently, the second term on the right-hand side (RHS) of \eqref{channel_estimate_eve_post} exists due to the use of correlated pilots assigned to different Bobs in the network. We highlight that the third term on the RHS of  \eqref{channel_estimate_eve_post} exists due to Eves' PC attack, which degrades the quality of channel estimates obtained by $\textrm{Alice}_l$. Since $N=0$ indicates no PC attack, the third term on the RHS of  \eqref{channel_estimate_eve_post} disappears when $N=0$. When $N\neq0$, the third term is always positive and PC attack reduces the channel estimation accuracy.

We note from \eqref{channel_estimate_eve_post} that the increase in PC depends on the strength of the interference caused by $\textrm{Eve}_{in}$, which can be controlled through adjusting $\bar{p}_{in}$, $\bar{\beta}_{in}$, and $\rho_{inlk}$. However, increasing the pilot transmit power, i.e., $\bar{p}_{in}$, also increases the chances that the PC attack is detected by $\textrm{Alice}_l$. Additionally, we highlight that $\bar{\beta}_{in}$ depends on the distance between $\textrm{Alice}_l$ and $\textrm{Eve}_{in}$. To increase $\bar{\beta}_{in}$, $\textrm{Eve}_{in}$ needs to move closer to $\textrm{Alice}_l$, which may not be possible due to mobility restrictions for $\textrm{Eve}_{in}$, as moving closer to $\textrm{Alice}_l$ will make $\textrm{Eve}_{in}$ more prone to detection. We clarify that by carefully selecting the pilot sequence for the PC attack, $\textrm{Eve}_{in}$ can increase $\rho_{inlk}$ without compromising its own privacy. We will discuss the selection of pilot sequences for the PC attack in Section~\ref{Sec:PCatt}.

\subsection{Downlink Transmission and SINR Under Pilot Contamination Attack}

In this subsection, we determine the downlink achievable SINR at $\textrm{Bob}_{lk}$ in the downlink transmission phase where $\textrm{Alice}_l$ sends data symbols to Bobs in the same cell. Based on the assumption of channel reciprocity in the TDD mode, $\textrm{Alice}_l$ uses the uplink channel estimates obtained via the uplink channel training, i.e., the channel estimates contaminated by Eves given by \eqref{channel_estimate_eve_post}, for the downlink transmission. We denote the symbols intended for $\textrm{Bob}_{lk}$ by $x_{lk}$ and assume that $\textrm{Alice}_l$ transmits $x_{lk}$ using the transmit power $P_{lk}$, where $P_{lk}=\mathbb{E}\left[x_{lk}^Hx_{lk}\right]$. Assuming that $\textrm{Alice}_l$ performs maximum-ratio-transmission (MRT) using the channel estimates contaminated by Eves, the precoding vector $\bar{\mathbf{t}}_{lk}$ for $\textrm{Bob}_{lk}$ under PC attack is given by
\begin{align}\label{precoding_vec_eve}
\bar{\mathbf{t}}_{lk}=\frac{1}{\sqrt{M{\bar{\alpha}}_{lk}}}\hat{\bar{\mathbf{h}}}_{lkl},
\end{align}
where
\begin{align}\label{alpha_corr_eve}
\bar{\alpha}_{lk}=\sum_{i=1}^{L}\sum_{j=1}^{K}p_{ij}\beta_{ijl}\left|\rho_{ijlk}\right|^{2}
+\sum_{i=1}^{L}\sum_{n=1}^N\bar{p}_{in}\bar{\beta}_{inl}\left|\rho_{inlk}\right|^{2}+\sigma_{z}^{2}.
\end{align}
We note that the second term on the RHS of \eqref{alpha_corr_eve} is caused by Eves' PC attack. Thus, Eves modify the precoder for $\textrm{Bob}_{lk}$ by contaminating the channel estimates such that the downlink achievable SINR for $\textrm{Bob}_{lk}$ is lower than that without the PC attack.

Next, $\textrm{Alice}_l$ multiplies $x_{lk}$ with $\bar{\mathbf{t}}_{lk}$ for the downlink transmission. Therefore, the downlink signal received at $\textrm{Bob}_{lk}$ is written as
\begin{align}\label{rec_initial}
\hat{\bar{r}}_{lk}=\sum_{i=1}^{L}\sum_{j=1}^{K}\sqrt{\beta_{lki}}
{\mathbf{h}_{lki}^H}\left(\bar{\mathbf{t}}_{ij}x_{ij}\right)+w,
\end{align}
where $w$ is the AWGN at $\textrm{Bob}_{lk}$. Assuming that $\textrm{Bob}_{lk}$ only has statistical information about its channel $\mathbf{h}_{lkl}$, we express the downlink signal received at $\textrm{Bob}_{lk}$ as
\begin{align}\label{received_sig_eve}
\hat{\bar{r}}_{lk}=&\sqrt{\beta_{lkl}} \mathbb{E} \left[\mathbf{h}_{lkl}^H\bar{\mathbf{t}}_{lk}\right]x_{lk} + \sqrt{\beta_{lkl}}\left(\mathbf{h}_{lkl}^H\bar{\mathbf{t}}_{lk}-\mathbb{E} \left[\mathbf{h}_{lkl}^H\bar{\mathbf{t}}_{lk}\right]\right)x_{lk}\notag\\
&+\sum_{(i,j)\neq (l,k)}\sqrt{\beta_{lki}}\mathbf{h}_{lki}^{H}\left(\bar{\mathbf{t}}_{ij}x_{ij}\right)+w.
\end{align}
Noting that the first term on the RHS of \eqref{received_sig_eve} is independent and uncorrelated with the remaining terms, the downlink SINR at $\textrm{Bob}_{lk}$ is written as
\begin{align}\label{long_exp_eve}
\bar{\theta}_{lk,M}=\frac{\left|\mathbb{E} \left[\mathbf{h}_{lkl}^H\bar{\mathbf{t}}_{lk}\right]\right|^2\beta_{lkl}P_{lk}}
{\text{var}\left[\mathbf{h}_{lkl}^H\bar{\mathbf{t}}_{lk}\right]\beta_{lkl}P_{lk}
+ \zeta_{lki}+ \sigma_{w}^2},
\end{align}
where we define $\zeta_{lki} = \sum_{(i,j)\neq (l,k)}\mathbb{E}[|\mathbf{h}_{lki}^H\bar{\mathbf{t}}_{ij}|^2]\beta_{lki}P_{ij}$. Afterwards, \eqref{long_exp_eve} can be written in closed-from as \cite{Akbar2016,Shen2015}
\begin{align}\label{SINR_eve}
\bar{\theta}_{lk,M}=\frac{\beta_{lkl}P_{lk}}{\bar{\alpha}_{lk}\left(\sum_{(i,j)\neq (l,k)} \frac{\left|\rho_{lkij}\right|^{2} \Xi_{lki}^2\beta_{lki}P_{ij}}{\bar{\alpha}_{ij}}\right)
+\frac{\bar{\alpha}_{lk}}{M}\left(P_{\textrm{tot}}\right)},
\end{align}
where $\Xi_{lki}^2 = {p_{lk}}\beta_{lki}$ and $P_{\textrm{tot}}=\sum_{i=1}^{L}\sum_{j=1}^{K}\beta_{lki}P_{ij}+\sigma_{w}^2$.

We highlight that \eqref{SINR_eve} is valid for an arbitrary number of antennas at the BS and an arbitrary number of Eves in the network. Notably, \eqref{SINR_eve} is also valid for the network having no Eve, i.e., $N=0$. For this network, we obtain the downlink achievable SINR at $\textrm{Bob}_{lk}$ as
\begin{align}\label{SINR}
{\theta}_{lk,M}=\frac{\beta_{lkl}P_{lk}}{{\alpha}_{lk}\left(\sum_{(i,j)\neq (l,k)} \frac{\left|\rho_{lkij}\right|^{2} \Xi_{lki}^2\beta_{lki}P_{ij}}{{\alpha}_{ij}}\right)
+\frac{{\alpha}_{lk}}{M}\left(P_{\textrm{tot}}\right)},
\end{align}
where
\begin{align}\label{alpha_corr}
{\alpha}_{lk}=\sum_{i=1}^{L}\sum_{j=1}^{K}p_{ij}\beta_{ijl}\left|\rho_{ijlk}\right|^{2}+\sigma_{z}^{2}.
\end{align}
Comparing \eqref{alpha_corr_eve} with \eqref{alpha_corr}, we note that utilizing the precoding vector contaminated by Eves results in an increased interference at $\textrm{Bob}_{lk}$ during the downlink data transmission phase.
\begin{remark}
By comparing \eqref{SINR_eve} with \eqref{SINR}, we observe that Eves are able to degrade the SINR at $\textrm{Bob}_{lk}$ with the PC attack. Importantly, the correlation parameter with Eves, given by \eqref{alpha_corr_eve}, is greater than correlation parameter without Eve, given by \eqref{alpha_corr}, i.e., $\bar{\alpha}_{lk} > \alpha_{lk}$. As a result, it is expected that the downlink SINR with Eves is smaller than that without Eves, i.e., $\bar{\theta}_{lk,M}<\theta_{lk,M}$.
\end{remark}

\section{User Capacity in Massive MIMO Networks Under Pilot Contamination Attack}\label{Sec:UCPC}

In this section, we examine the impact of the PC attack on the user capacity of the considered massive MIMO network. Here, we define the user capacity as the bound on the number of Bobs that can be simultaneously served via the downlink of the network such that the SINR requirement for all Bobs can be satisfied.

We assume that the SINR requirement for $\textrm{Bob}_{lk}$ is $\gamma_{lk}$. Also, we assume that the downlink achievable SINR at $\textrm{Bob}_{lk}$ needs to be higher than the SINR requirement $\gamma_{lk}$ for successful downlink transmission, i.e, $\bar{\theta}_{lk,M}\geq\gamma_{lk}$ for $M$ antennas at $\textrm{Alice}_l$. We present the bound on the user capacity of the network in the following proposition.
\begin{proposition}\label{prop_K_tot}
For a multi-cell multiuser massive MIMO network under PC attack, where the transmit power of $\textrm{Eve}_{in}$ is $\bar{p}_{in}>0$ during the uplink transmission phase, the bound on the total number of Bobs, $K_\textrm{tot}$, that can be simultaneously served via the downlink transmission is given by
\begin{align}\label{K_to_eve}
K_{\textrm{tot}}\leq\left[\left(\frac{\tau}{LN+1}\right)\sum_{l=1}^{L}\sum_{k=1}^{K}\left(\frac{1+\gamma_{lk}}{\gamma_{lk}}\right)\right]^{\frac{1}{2}},
\end{align}
where $K_\textrm{tot}=LK$ and $\frac{\gamma_{lk}}{1+\gamma_{lk}}$ is the effective bandwidth of $\textrm{Bob}_{lk}$.
\begin{IEEEproof}
The proof is presented in Appendix~\ref{appa}.
\end{IEEEproof}
\end{proposition}

We note that the bound on the user capacity given in \eqref{K_to_eve} is a generalized expression and valid for any pilot sequence design. However, not every pilot design is capable of achieving the bound in \eqref{K_to_eve} with equality \cite{Akbar2016,Shen2015}. For most pilot designs, \eqref{K_to_eve} is satisfied without equality, i.e., $K_{\textrm{tot}}<\left[\left(\frac{\tau}{LN+1}\right)\sum_{l=1}^{L} \sum_{k=1}^{K}\left(\frac{1+\gamma_{lk}}{\gamma_{lk}}\right)\right]^{\frac{1}{2}}$. As such, different pilot designs are capable of supporting different numbers of Bobs in the network. We highlight that a carefully designed pilot sequence, such as the use capacity-achieving pilot design \cite{Akbar2016,Shen2015}, is capable of achieving the bound on the user capacity with equality, i.e., $K_{\textrm{tot}}=\left[\left(\frac{\tau}{LN+1}\right) \sum_{l=1}^{L}\sum_{k=1}^{K}\left(\frac{1+\gamma_{lk}}{\gamma_{lk}}\right)\right]^{\frac{1}{2}}$. As such, the user capacity-achieving pilot design can satisfy a diverse range of SINR requirements for various Bobs in the network.

\begin{table*}[!t]
\centering
\begin{minipage}{\textwidth}
\renewcommand{\arraystretch}{2.5}
\caption{List of Important Formulas}
\label{List of Formulas}
\resizebox{\textwidth}{!}{\begin{tabular}{|l|l|l|} \hline
 & \textbf{Without PC Attack $N=0$} & \textbf{With PC Attack $N>0$} \\ \hline

\textbf{Channel Estimates} & $\hat{{\mathbf{h}}}_{lkl}=\sum_{i=1}^L \sum_{j=1}^{K}\sqrt{p_{ij}\beta_{ijl}}\rho_{ijlk}\mathbf{h}_{ijl} +\mathbf{z}$ &
$\hat{\bar{\mathbf{h}}}_{lkl}=\sum_{i=1}^L \sum_{j=1}^{K}\sqrt{p_{ij}\beta_{ijl}}\rho_{ijlk}\mathbf{h}_{ijl}
+\sum_{i=1}^{L}\sum_{n=1}^N\sqrt{\bar{p}_{in}\bar{\beta}_{inl}}\rho_{inlk}\bar{\mathbf{h}}_{inl}+\mathbf{z}~~(2)$ \\ \hline

\textbf{Correlation Parameter} & ${\alpha}_{lk}=\sum_{i=1}^{L}\sum_{j=1}^{K}p_{ij}\beta_{ijl}\left|\rho_{ijlk}\right|^{2}+\sigma_{z}^{2}$ & $\bar{\alpha}_{lk}=\sum_{i=1}^{L}\sum_{j=1}^{K}p_{ij}\beta_{ijl}\left|\rho_{ijlk}\right|^{2}
+\sum_{i=1}^{L}\sum_{n=1}^N\bar{p}_{in}\bar{\beta}_{inl}\left|\rho_{inlk}\right|^{2}+\sigma_{z}^{2}~~(4)$ \\ \hline

\textbf{Precoding Vector} & $\mathbf{t}_{lk}=\frac{1}{\sqrt{M\alpha_{lk}}}\hat{\mathbf{h}}_{lkl}$ & $\bar{\mathbf{t}}_{lk}=\frac{1}{\sqrt{M{\bar{\alpha}}_{lk}}}\hat{\bar{\mathbf{h}}}_{lkl}~~(3)$ \\ \hline

\textbf{Downlink SINR} & ${\theta}_{lk,M}=\frac{\left|\mathbb{E} \left[\mathbf{h}_{lkl}^H{\mathbf{t}}_{lk}\right]\right|^2\beta_{lkl}P_{lk}}
{\text{var}\left[\mathbf{h}_{lkl}^H{\mathbf{t}}_{lk}\right]\beta_{lkl}P_{lk}
+ \sum_{(i,j)\neq (l,k)}\mathbb{E}[|\mathbf{h}_{lki}^H{\mathbf{t}}_{ij}|^2]\beta_{lki}P_{ij}+ \sigma_{w}^2}$ & $\bar{\theta}_{lk,M}=\frac{\left|\mathbb{E} \left[\mathbf{h}_{lkl}^H\bar{\mathbf{t}}_{lk}\right]\right|^2\beta_{lkl}P_{lk}}
{\text{var}\left[\mathbf{h}_{lkl}^H\bar{\mathbf{t}}_{lk}\right]\beta_{lkl}P_{lk}
+ \sum_{(i,j)\neq (l,k)}\mathbb{E}[|\mathbf{h}_{lki}^H\bar{\mathbf{t}}_{ij}|^2]\beta_{lki}P_{ij}+ \sigma_{w}^2}~~(7)$  \\ \hline

\textbf{Downlink SINR (closed-form)} & ${\theta}_{lk,M}=\frac{\beta_{lkl}P_{lk}}{{\alpha}_{lk}\left(\sum_{(i,j)\neq (l,k)}^K \frac{\left|\rho_{lkij}\right|^{2} \Xi_{lki}^2\beta_{lki}P_{ij}}{{\alpha}_{ij}}\right)
+\frac{{\alpha}_{lk}}{M}\left(\sum_{i=1}^{L}\sum_{j=1}^{K}\beta_{lki}P_{ij}+\sigma_{w}^2\right)}$ & $\bar{\theta}_{lk,M}=\frac{\beta_{lkl}P_{lk}}{\bar{\alpha}_{lk}\left(\sum_{(i,j)\neq (l,k)}^K \frac{\left|\rho_{lkij}\right|^{2} \Xi_{lki}^2\beta_{lki}P_{ij}}{\bar{\alpha}_{ij}}\right)
+\frac{\bar{\alpha}_{lk}}{M}\left(\sum_{i=1}^{L}\sum_{j=1}^{K}\beta_{lki}P_{ij}+\sigma_{w}^2\right)}~~(9)$ \\ \hline

\textbf{Downlink User Capacity Region} & $\sum_{l=1}^{L}\sum_{k=1}^{K}\frac{\gamma_{lk}}{1+\gamma_{lk}}\leq \tau$ & $\sum_{l=1}^{L}\sum_{k=1}^{K}\frac{\gamma_{lk}}{1+\gamma_{lk}}\leq \frac{\tau}{LN+1}~~(12)$ \\ \hline
\end{tabular}}
\end{minipage}
\end{table*}

\begin{proposition}\label{prop_K_tot_new}
The bound on the user capacity given by \eqref{K_to_eve} with $N$ Eves in each cell of the network holds when the sum of effective bandwidths of all Bobs in the network is less than or equal to the ratio between the length of the pilot sequence and $(LN+1)$. We express this condition as
\begin{align}\label{BW_all_eve}
\sum_{l=1}^{L}\sum_{k=1}^{K}\frac{\gamma_{lk}}{1+\gamma_{lk}}\leq \frac{\tau}{LN+1}.
\end{align}
\begin{IEEEproof}
Utilizing the Cauchy-Schwarz inequality, we obtain
\begin{align}\label{proof_prop2}
\sum_{l=1}^{L}\sum_{k=1}^{K}\frac{1+\gamma_{lk}}{\gamma_{lk}}\geq\frac{K_{\textrm{tot}}^2}{\sum_{l=1}^{L} \sum_{k=1}^{K}\frac{\gamma_{lk}}{1+\gamma_{lk}}}.
\end{align}
Using \eqref{BW_all_eve}, we simplify \eqref{proof_prop2} as
\begin{align}\label{proof_prop21}
\sum_{l=1}^{L}\sum_{k=1}^{K}\frac{1+\gamma_{lk}}{\gamma_{lk}}\geq \frac{K_{\textrm{tot}}^2\left(LN+1\right)}{\tau}.
\end{align}
We next simplify the expression \eqref{proof_prop21} to obtain \eqref{K_to_eve}, which completes the proof.
\end{IEEEproof}
\end{proposition}

From \eqref{K_to_eve}, we note that the bound on the user capacity of the network without Eves is a special case of \eqref{K_to_eve} when $N=0$. Specifically, the bound on the user capacity region of the massive MIMO network without Eves is obtained as
\begin{align}\label{K_tot}
K_{\textrm{tot}}\leq\left(\tau\sum_{l=1}^{L}\sum\limits_{k=1}^{K}\frac{1+\gamma_{lk}}{\gamma_{lk}}\right)^{\frac{1}{2}}.
\end{align}
Furthermore, the user capacity given in \eqref{K_tot} is satisfied when the sum of the effective bandwidths of all Bobs in the network are less than or equal to the length of the pilot sequence \cite{Akbar2016}. This condition is expressed as
\begin{align}\label{BW_all}
\sum_{l=1}^{L}\sum_{k=1}^{K}\frac{\gamma_{lk}}{1+\gamma_{lk}}\leq\tau.
\end{align}

We note that the bound in \eqref{BW_all} specifies the user capacity region of the network without Eves. Notably, the bound on the user capacity in \eqref{K_tot} is satisfied as long as the bound on the user capacity region in \eqref{BW_all} holds. For a given user capacity and user capacity region, it is possible to design pilot sequences for all Bobs in the network such that the Bobs' SINR requirements are satisfied with a limited number of antennas at Alice \cite{Akbar2016,Akbar2016b,Shen2015}. We refer to this pilot sequence design as the \emph{user capacity-achieving pilot sequence design}. We further refer to the designed pilot sequences as \emph{user capacity-achieving pilot sequences}.

\begin{remark}
Comparing \eqref{BW_all_eve} and \eqref{BW_all}, we note that the user capacity region is significantly reduced in the presence of the PC attack. The reduction in the user capacity region depends on the number of Eves, $N$, in the network.
\end{remark}

Table~\ref{List of Formulas} provides a summary of important formulas in Section.~\ref{Sec:System_Model} and Section.~\ref{Sec:UCPC}. We highlight that the network without the PC attack is a special case of a network with the PC attack for $N=0$.

\section{Capacity-Achieving Pilot Sequence Design and Power Allocation}\label{Sec:Design}

In this section, we present the algorithm for the user capacity-achieving pilot sequence design. The algorithm designs pilot sequences such that the SINR requirements for all Bobs in the network are satisfied. We assume that the pilot sequences are designed by $\textrm{Alice}_l$ based on the individual SINR requirements for Bobs in the same cell. These pilot sequences are then used by Bobs during the uplink channel estimation phase. We assume that each Bob has a minimum SINR requirement for the downlink transmission, i.e., $\textrm{Bob}_{lk}$ has the minimum SINR requirement of $\gamma_{lk}$. The goal of the user  capacity-achieving pilot design is to generate correlated pilots such that
$\theta_{lk,M}$ is greater than or equal to $\gamma_{lk}$. We next detail the pilot sequence design process.

\subsection{Pilot Sequence Design at $\textrm{Alice}$}

\subsubsection{Computation of Effective Bandwidths}

We assume that $\textrm{Alice}_l$ knows the SINR requirements for all Bobs in the $l$-th cell. We represent the SINR requirement for all Bobs in the $l$-th cell by a vector $\mathbf{b} = [\gamma_{l1},\gamma_{l2}, \dotsc, \gamma_{lK}]$, where $\gamma_{l1}\geq\gamma_{l2}\geq\cdots\geq\gamma_{lK}$. Based on the information of $\mathbf{b}$, Alice calculates the effective bandwidths for all Bobs as $\bar{\mathbf{b}}=[\gamma_{l1}/(1+\gamma_{l1}),\gamma_{l2}/(1+\gamma_{l2}),\cdots,\gamma_{lK}/(1+\gamma_{lK})]$. We note that the SINR requirements must be chosen to achieve the bound on the user capacity region given in \eqref{BW_all}. We highlight that it is desirable to achieve \eqref{BW_all} with equality such that the benefits of a lager user capacity region are fully utilized \cite{Akbar2016}. As such, Alice modifies $\bar{\mathbf{b}}$ as $\hat{\mathbf{b}}=[\hat{b}_{1},\hat{b}_{2},\cdots,\hat{b}_{K}]$, where $\hat{b}_{k}=\hat{\gamma}_{lk}/(1+\hat{\gamma}_{lk})$ and $\sum_{k=1}^{K}\hat{b}_{k}=\tau$.
We highlight that the SINR modification needs to be carefully carried out such that they remain inside the user capacity region.

\subsubsection{Majorization and T-Transform}

In this step, $\textrm{Alice}_l$ aims to find a vector $\mathbf{x}=\left[x_{1},\cdots,x_{\tau},0,\cdots,0\right]$ that majorizes $\hat{\mathbf{b}}$, i.e., $\mathbf{x}\succ\hat{\mathbf{b}}$. We highlight that $\mathbf{x}$ is easily obtained from $\hat{\mathbf{b}}$ when $x_{p}=\frac{1}{\tau}\sum_{k=1}^{K}\hat{b}_{k}=1$, where $p\in\left\{1,\cdots,\tau\right\}$ and $x_{q}=0$, where $q\in\left\{\tau+1,\cdots,K\right\}$. Given $\mathbf{x}$,
$\hat{\mathbf{b}}$ is obtainable by applying at most $K-1$ T-transform operations \cite{Hardy1954} on $\mathbf{x}$, i.e., $\hat{\mathbf{b}}=\mathbf{T}_{K-1}\mathbf{T}_{K-2}\cdots\mathbf{T}_{1}\mathbf{x}$, and there exists a matrix $\mathbf{W}=\mathbf{W}_{1}\mathbf{W}_{2}\cdots\mathbf{W}_{K-1}$, where $\mathbf{W}_{k}$ is a unitary matrix, which is generated from $\mathbf{T}_{k}$ at each step of the T-transform \cite{Hardy1954,Viswanath1999,Akbar2016}.

\subsubsection{Generation of User Capacity-Achieving Pilot Sequences}

In this step, $\textrm{Alice}_l$ generates pilot sequences for all Bobs in the $l$-th cell based on $\mathbf{x}$, $\hat{\mathbf{b}}$ and $\mathbf{W}$. $\textrm{Alice}_l$ obtains a matrix $\mathbf{V}$ from $\mathbf{W}$, where $\mathbf{V}$ contains only the first $\tau$ rows from $\mathbf{W}$. Afterwards, $\textrm{Alice}_l$ obtains the pilot sequence for all Bobs in the $l$-th cell as
\begin{align}\label{K_tos}
\mathbf{S}_l=\text{normc}\left(B^{\frac{1}{2}}\mathbf{V}\mathbf{Z}^{-\frac{1}{2}}\right),
\end{align}
where $B=\sum_{k=1}^{K}\hat{b}_{k}$ and $\mathbf{Z}=\textrm{diag}\left(\hat{\mathbf{b}}\right)$. We note that each column of $\mathbf{S}_l$ represents the pilot sequence vector for one Bob in the $l$-th cell. Specifically, $\mathbf{S}_l = [\mathbf{s}_{l1},\cdots,\mathbf{s}_{lk},\cdots,\mathbf{s}_{lK}]$. We highlight that the user capacity-achieving pilot sequence design method is based on the principles of generalized Welch bound equality sequences used in code-division multiple access systems \cite{Welch1974}.

\subsection{Downlink Power Allocation at Alice}

During the downlink data transmission phase, $\textrm{Alice}_l$ carefully chooses the downlink transmission power with the aim to satisfy the SINR requirers for all Bobs in the $l$-th cell. Specifically, $\textrm{Alice}_l$ allocates the downlink power for $\textrm{Bob}_{lk}$ as \cite{Akbar2016,Shen2015}
\begin{align}\label{dp_alloc}
P_{lk}=c\frac{\hat{\gamma}_{lk}}{1+\hat{\gamma}_{lk}},
\end{align}
where $c>0$ is a constant power-scaling factor. Moreover, \cite{Akbar2016} proposed a specific power allocation scheme where the power-scaling factor for $P_{lk}$ is chosen as $c=\alpha_{lk}$. When $c=\alpha_{lk}$ is used for $\textrm{Bob}_{lk}$, the user capacity-achieving pilot sequence design satisfies the SINR requirements of all the Bobs in the network, where there is no pilot contamination attack, i.e., $N=0$ \cite{Akbar2016}. We highlight that when the power allocation scheme is used with the user capacity-achieving pilot sequence design, it is possible to achieve the SINR requirements for all Bobs in the network provided that the SINR requirements lie inside the user capacity region \cite{Akbar2016,Shen2015}.
\begin{remark}
We note that $\textrm{Alice}_l$ can design a valid pilot sequence set as long as the SINR requirements lie inside the user capacity region. However, when $N$ is large, the area under the user capacity region is small. Consequently, even if $\textrm{Alice}_l$ is aware of the PC attack, it may not be possible for $\textrm{Alice}_l$ to design pilot sequences to satisfy the SINR requirements for all Bobs in the $l$-th cell when $N$ is large.
\end{remark}

\section{Vulnerabilities in User Capacity-Achieving Pilot Sequence Design}\label{Sec:Vul}

In this section, we identify the potential vulnerabilities in the user capacity-achieving pilot sequence design. Specifically, we highlight that these vulnerabilities make the massive MIMO network prone to PC attack.

\subsection{Structure of Pilot Sequence Set}\label{mathrefs}

We note that the pilot sequence for at least one Bob in each cell is the same, regardless of the SINR requirements for all Bobs in the network. This limitation stems from the design of the user capacity-achieving pilot method. We highlight that the pilot sequence for at least one Bob in each cell is $[1, 0, \cdots, 0, 0]^T$, where the pilot sequence is a column vector with length $\tau$. As such, based on the knowledge that the user capacity-achieving pilot design is used in the network, Eves knows the pilot sequence for at least one Bob per cell. Consequently, it is logical for Eves to transmit $[1, 0, \cdots, 0, 0]^T$ for the PC attack in the network where the pilot sequences are obtained by using the user capacity-achieving pilot sequence design.

\subsection{SINR Modification in User Capacity-Achieving Pilot Design}

We highlight that the user capacity-achieving pilot sequence design modifies the SINR requirements for all Bobs by choosing $\hat{\gamma}_{lk}\geq\gamma_{lk}$. The SINR modification is performed such that the SINR requirements lie on the upper surface boundary of the user capacity region, i.e, $\sum_{l=1}^{L}\sum_{k=1}^{K}\frac{\hat{\gamma}_{lk}}{1+\hat{\gamma}_{lk}}=\frac{\tau}{LN+1}$. We note that this SINR modification in the user capacity-achieving pilot sequence design makes the network more prone to PC attack. Moreover, it is possible that the SINR requirements may lie outside the user capacity region in the presence of the PC attack. We highlight that the user capacity-achieving pilot sequence design is undefined for the area outside the user capacity region. As such, the pilot sequence design no longer guarantees that the SINR requirements for all the users are satisfied when they lie outside the user capacity region. Furthermore, we observe from \eqref{BW_all_eve} that the user capacity region is significantly reduced in the presence of even a single active attacker. Importantly, the SINR requirements may lie outside the user capacity region in a network under the PC attack, which undermines the benefits of the user capacity-achieving pilot sequence design.

\subsection{Correlated Pilots}

The pilot sequences designed by the user capacity-achieving pilot sequence design are correlated, i.e., $\rho_{ijlk}\neq0$. The user capacity-achieving pilot sequence design aim at controlling the cross-correlation between pilot sequences to manage PC. However, due to the use of correlated pilots, the PC attack by using the known pilot sequences identified in Sec.~\ref{mathrefs} can potentially contaminate the channel estimates for Bobs with the pilot sequences unknown to Eves.

\section{Pilot Contamination Attack Strategy Adopted by Eves}\label{Sec:PCatt}

In this section, we outline a simple yet effective PC attack strategy. We assume that Eves has limited knowledge about the network. However, we note that Eves need to know two network parameters for successfully exploiting the user capacity-achieving pilot sequence design, i.e., the length of pilot sequence and the information that user capacity-achieving pilot design is used in the network. We highlight that these parameters are easy to obtain in any network. Throughout this paper, we assume that Eves have knowledge of these network parameters before the uplink channel training phase.

\subsection{Attacking Aims of Eves}

As active attackers, Eves aim to disturb the functioning of the network. Specifically, Eves aims to:
\begin{enumerate}
\item
Exploit the vulnerabilities in the user capacity-achieving pilot sequence design to increase PC in the uplink channel estimates;
\item
Degrade the achievable SINR of the user with known pilot sequence, i.e., $\textrm{Bob}_{lk}$, such that the SINR target for $\textrm{Bob}_{lk}$ cannot be satisfied even with an unlimited number of antennas at $\textrm{Alice}_l$;
\item
Deteriorate the achievable SINR for users with unknown pilot sequence such that their respective SINR targets are no longer satisfied with the pre-determined number of antennas at the $\textrm{Alice}_l$.
\end{enumerate}
We recall from \eqref{rec_pilot_eve} that the $\textrm{Eve}_{in}$ has some degree of control over three parameters, i.e., $\bar{p}_{in}$, $\bar{\beta}_{inl}$, and $\bar{\mathbf{S}}_{in}$. We next detail each parameter in the following subsections.

\subsection{Transmit Power $\bar{p}_{in}$ at $\textrm{Eve}_{in}$ During Uplink Training}

During the uplink channel estimation phase, $\textrm{Eve}_{in}$ transmits the pilot sequence with the transmit power $\bar{p}_{in}$. Furthermore, $\textrm{Eve}_{in}$ can transmit with a small non-zero transmit power to avoid being detected. We note that $\textrm{Eve}_{in}$ can increase PC in the channel estimates at $\textrm{Alice}_l$ by increasing $\bar{p}_{in}$. We also note that the increase in $\bar{p}_{in}$ increases the probability of detection of $\textrm{Eve}_{in}$ by $\textrm{Alice}_l$. As such, we make a reasonable assumption that $\bar{p}_{in}$ cannot be greater than the highest transmit power amongst all Bobs in the $l$-th cell in the network network during the uplink channel estimation phase. We represent this condition as
\begin{align}\label{pcpower}
0<\bar{p}_{in}\leq\textrm{max}\left\{p_{lk}\right\},~\textrm{where}~k\in\{1,\dotsc,K\}.
\end{align}

\subsection{Locations of Eves}

We note that the location of $\textrm{Eve}_{in}$ is important in the success of the PC attack. We highlight that the large-scale propagation coefficient $\bar{\beta}_{inl}$ is a distance-dependent parameter. As such, an Eve can locate itself close to the BS to increase PC. However, the mobility of Eves may not be practical in order to avoid detection by the BS. Additionally, the BS may perform a simple inverse-power control to mitigate the effect of the large-scale propagation coefficient. The inverse-power control is applied such that $p_{lk}\beta_{lkl}=1$. As such, there is an additional constraint on the pilot transmit power of $\textrm{Eve}_{in}$, given by
\begin{align}\label{pcpower2}
\bar{p}_{in}\bar{\beta}_{inl}\leq 1.
\end{align}
This condition ensures that the PC power from $\textrm{Eve}_{in}$ received at $\textrm{Alice}_l$ does not exceed the power received from Bobs during the uplink channel estimation phase.

\subsection{Pilot Sequence Transmitted by Eves}
In this subsection, we present the optimal pilot sequence adopted at $\textrm{Eve}_{in}$ for the PC attack on $\textrm{Bob}_{lk}$. Moreover, we examine the impact of Eves' PC attack on the uplink channel estimates of Bobs whose pilot sequences are not known to Eves.
\begin{proposition}\label{prop_mse}
For a multi-cell multiuser massive MIMO network under PC attack, the optimal pilot sequences transmitted by $\textrm{Eve}_{in}$ to maximize the error variance between the channel estimate for $\textrm{Bob}_{lk}$ without the PC attack and the channel estimate for $\textrm{Bob}_{lk}$ with the PC attack are given by $\bar{\mathbf{s}}_{in}={\mathbf{s}}_{lk}$ and $\bar{\mathbf{s}}_{in}={\mathbf{s}}_{lk}^{\ast}$, respectively.
\begin{IEEEproof}
From \eqref{channel_estimate_eve_post}, we obtain the error between the channel estimate for $\textrm{Bob}_{lk}$ with no PC and the channel estimate for $\textrm{Bob}_{lk}$ with the PC attack as
\begin{align}\label{mse1}
\xi_{lk}=\hat{\bar{\mathbf{h}}}_{lkl}-\hat{\mathbf{h}}_{lkl} =\sum_{i=1}^{L}\sum_{n=1}^N\sqrt{\bar{p}_{in}\bar{\beta}_{inl}}\rho_{inlk}\bar{\mathbf{h}}_{inl},
\end{align}
where $\hat{\mathbf{h}}_{lkl}$ is the channel estimate for $\textrm{Bob}_{lk}$ without the PC attack and is obtained from \eqref{channel_estimate_eve_post} when $N=0$. We next obtain the variance of $\xi_{lk}$ from \eqref{mse1} as
\begin{align}\label{mse2}
\textrm{var}\left(\xi_{lk}\right)&=\mathbb{E}\left[\xi_{lk}^{H}\xi_{lk}\right],\notag\\
\hspace{2mm} &=\mathbb{E}\Bigg[\left(\sum_{i=1}^{L}\sum_{n=1}^N \sqrt{\bar{p}_{in} \bar{\beta}_{inl}} \rho_{inlk}\bar{\mathbf{h}}_{inl}^H\right) \times \\
& \hspace{0.9cm} \left(\sum_{p=1}^{L}\sum_{q=1}^N \sqrt{\bar{p}_{pq} \bar{\beta}_{pql}} \rho_{pqlk}\bar{\mathbf{h}}_{pql}\right)\Bigg],\notag\\
\hspace{2mm}&=\sum_{i=1}^L \sum_{n=1}^N \bar{p}_{in} \bar{\beta}_{inl} \rho_{inlk}\mathbb{E}\left[\bar{\mathbf{h}}_{inl}^H\bar{\mathbf{h}}_{inl}\right].
\end{align}
We next utilize the channel hardening property of massive MIMO, which is given by \cite{Akbar2016,Shen2015}
\begin{align}\label{ortho_mm}
\lim_{M \rightarrow \infty} \frac{1}{M}\mathbf{h}_{ijl}^{H}\mathbf{h}_{lkl}=
\begin{cases}
1, & \forall~(i,j)=(l,k)\\
0, & \text{otherwise,}
\end{cases}
\end{align}
to simplify \eqref{mse2} under the assumption that $M\rightarrow \infty$ as
\begin{align}\label{mse3}
\textrm{var}\left(\Xi_{lk}\right)=M\sum_{i=1}^{L}\sum_{n=1}^{N}\bar{p}_{in}\bar{\beta}_{inl}|\rho_{inlk}|^{2}.
\end{align}
Using \eqref{pcpower2} and \eqref{mse3}, we obtain
\begin{align}\label{mse4}
\textrm{var}\left(\Xi_{k}\right)\leq M \sum_{i=1}^{L}\sum_{n=1}^{N}|\rho_{inlk}|^{2}.
\end{align}
We recall that $\rho_{inlk}=\mathbf{s}_{lk}^{T}\bar{\mathbf{s}}_{in}$, where the value of $\rho_{inlk}$ is between $-1$ and $+1$ for user capacity achieving pilot sequence design. Also, we note that the maximum value of $\sum_{i=1}^{L}\sum_{n=1}^N|\rho_{inlk}|^2$ is $LN$, which is achieved when $\bar{\mathbf{s}}_{in}=\mathbf{s}_{lk}$ or $\bar{\mathbf{s}}_{in}=\mathbf{s}_{lk}^{\ast}$. From \eqref{mse4}, we note that the maximum value of $\textrm{var}\left(\Xi_{lk}\right)$ is $MLN$, when all Eves transmit the known pilot sequence. As such, $\textrm{var}\left(\Xi_{lk}\right)$ is maximum when all Eves transmit the same pilot sequence allocated to $\textrm{Bob}_{lk}$, which completes the proof.
\end{IEEEproof}
\end{proposition}

We highlight that maximizing error variance of the channel estimates also impacts the downlink precoder. Alternatively, it is possible to directly compute the pilot contamination precoder through eigenvalue analysis of the correlation matrix \cite{Wu2016}. We next examine the impact of Eves' PC attack using the known pilot sequences on Bobs with the unknown pilot sequences in the network in the following \textit{Lemma}.
\begin{lemma}
For a multi-cell multiuser massive MIMO network under PC attack, where a non-orthogonal pilot sequence set is used in the network, transmitting the known pilot sequence by Eves contaminates the channel estimates of Bobs with unknown pilot sequences.
\begin{IEEEproof}
We note that the channel estimates are contaminated when $\rho_{inlk}\neq0$. Notably, when a non-orthogonal pilot sequence set is used in the network, $\rho_{inlk}\neq0$ is achieved for all Bobs in the network. Specifically, pilot designs, e.g., the user capacity-achieving sequence design, have a non-zero cross-correlation for all Bobs. Consequently, we observe from \eqref{mse4} that when Eves transmit the known pilot sequence used by $\textrm{Bob}_{lk}$, we have $\textrm{var}\left(\Xi_{lk}\right)=MLN$. Furthermore, $\textrm{var}\left(\Xi_{ij}\right)>0$, where $(i,j)\neq (l,k)$ and $\rho_{ijlk}\neq 0$. As such, transmitting the known pilot sequences by Eves in a network with correlated pilot sequences not only degrades the performance of Bobs with known pilot sequences but degrades the performance of Bobs with unknown pilot sequences.
\end{IEEEproof}
\end{lemma}

\subsection{Detection and Mitigation of PC Attack}
In this subsection, we discuss the detection and mitigation of PC attack from the perspective of $\textrm{Alice}_l$. We highlight that $\textrm{Eve}_{in}$ utilizes the known vulnerabilities in the user capacity-achieving pilot design for PC attack. As such, the target user $\textrm{Bob}_{lk}$ archives an SINR lower than the target $\gamma_{lk}$. To detect the PC attack on $\textrm{Bob}_{lk}$, $\textrm{Alice}_l$ can request feedback from $\textrm{Bob}_{lk}$ on whether $\theta_{lk} \geq \gamma_{lk}$ occurs in the previous channel coherence interval. Based on the feedback from $\textrm{Bob}_{lk}$, multiple unsuccessful transmissions in the downlink for $\textrm{Bob}_{lk}$ indicate a possible PC attack on $\textrm{Bob}_{lk}$. After detecting the attack, $\textrm{Alice}_l$ can perform a reactive action and redesign the pilot sequences for the $l$-th cell. As such, $\textrm{Bob}_{lk}$ is assigned a pilot sequence that is different from the one known to $\textrm{Eve}_{in}$. This implies that $\textrm{Alice}_l$ can mitigate the impact of the PC attack on $\textrm{Bob}_{lk}$ by redesigning pilot sequences if the attack is detected.

Another possible method for $\textrm{Alice}_l$ to mitigate the impact of PC attack is to use a proactive approach in pilot sequence design. For example, $\textrm{Alice}_l$ may anticipate the PC attack and carefully design the pilot sequences for all Bobs in the $l$-th cell. Importantly, $\textrm{Alice}_l$ may sacrifice the advantages offered by a higher per-cell user capacity region by designing the pilot sequences according to
\begin{align}\label{BW_all22}
\sum_{l=1}^{L}\sum_{k=1}^{K}\frac{\gamma_{lk}}{1+\gamma_{lk}}\leq\tau - \delta_l
\end{align}
instead of \eqref{BW_all}, where $\delta_l$ is the factor indicating the anticipated reduction in the per-cell user capacity region in the $l$-th cell under the PC attack. Thus, the pilot sequences designed by $\textrm{Alice}_l$ using \eqref{BW_all22} lie inside the per-cell user capacity region. It follows that the SINR requirements for all the users in the $l$-th cell are satisfied, even in the presence of the PC attack. Furthermore, it is possible to detect and mitigate PC attack by including a random training phase after the uplink training phase \cite{Tian2017}.

\section{Numerical Results}\label{sec:num_results}

In this section, we numerically analyze the performance of the massive MIMO network under the PC attack. We first evaluate the impact of Eves on the user capacity region. Then, we examine the impact of Eves on the achievable SINR when $\textrm{Alice}_l$ is equipped with a large but limited number of antennas. A summary of important simulation parameters is given in Table~\ref{table_simulation}. Throughout this section, the simulation parameters remain the same unless we specifically state otherwise.

\begin{table}[!t]
\centering
\caption{A Summary of Simulation Parameters}
\label{table_simulation}
\centering
\begin{tabular}{|l|l|} \hline
\textbf{Parameter}        &  \textbf{Value} \\ \hline \hline
$L$                       &  2            \\ \hline
$K$                       &  4, except for Figs.~2, 3, and 5             \\ \hline
$\tau$                    &  3, except for Fig.~2              \\ \hline
SINR Requirements         &  Fig.~2: $[\gamma_{l1}, \gamma_{l2}, \gamma_{l3}, 1, 1]$ \\
                          &  Figs.~3 and 5: $[\gamma_{l1}=\gamma_{l2}=\dotsc=\gamma_{lK}=\gamma]$ \\
                          &  Fig.~4: $\bm{\gamma}_1 = \left[\gamma, \gamma, \gamma, \eta\gamma\right]$, $\bm{\gamma}_2=\left[\eta\gamma, \eta\gamma, \eta\gamma, \eta\gamma\right]$ \\
                          &  Figs.~6, 7, and 8: $\bm{\gamma}_1 = [0.91, 0.74, 0.64, 0.23]$ \\
                          & $\bm{\gamma}_2 = [0.94, 0.82, 0.45, 0.20]$ \\ \hline
$\bar{p}_{in}$                    &  0 dB, except for Fig.~6              \\ \hline
$\bar{p}_{lk}\beta_{lkl}$        &  1 (due to uplink power control)              \\ \hline
$P_{lk}$                       &  $\alpha_{lk}\left(\frac{\hat{\gamma}_{lk}}{1+\hat{\gamma}_{lk}}\right)$              \\ \hline
\end{tabular}
\end{table}

\subsection{Impact of PC Attack on User Capacity Region}

In this subsection, we examine the performance of the network with a sufficiently large number of antennas at $\textrm{Alice}_l$. The results in this subsection are obtained from \eqref{BW_all_eve}, which represents the bound on the user capacity of the massive MIMO network with $N$ Eves.

\begin{figure}[!t]
\centering
    \includegraphics[height=2.65in,width=0.95\columnwidth]{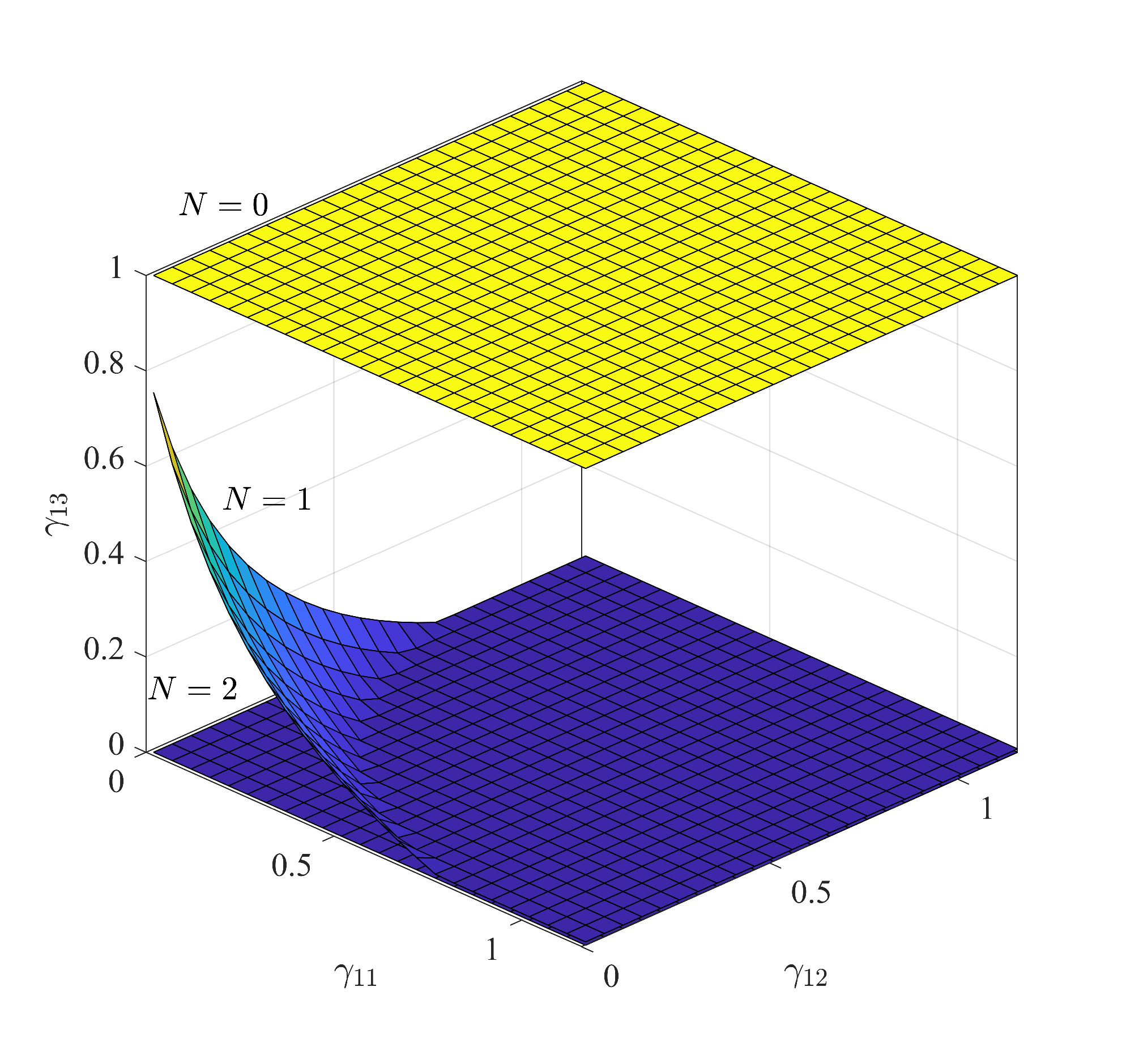}
    \caption{The reduction in the per-cell user capacity region in the massive MIMO network under the PC attack for different number of Eves in a cell.}\label{cap_region}
\end{figure}

We first demonstrate the reduction in the user capacity region when the network is under the PC attack. The SINR requirements for all Bobs in the massive MIMO network are given by the vector $[\gamma_{l1}, \gamma_{l2}, \gamma_{l3}, 1, 1]$. Furthermore, we assume $K=5$ and $\tau=3$. Fig. \ref{cap_region} shows the upper surface boundary of the per-cell user capacity region for different values of $N$. Here, $N=0$ indicates no Eves in the network. For the purpose of visualization, we set $\gamma_{13} = \min\{1,\gamma_{13}\}$. We note that the per-cell user capacity region is significantly reduced in the presence of Eves. For example, when $N=1$ and $N=2$, there is a $93\%$ and $99\%$ reduction in the area under the user capacity region, respectively, as compared to $N=0$. The reduction in the user capacity region signifies that the network is no longer capable of satisfying a diverse range of SINR requirements when the user capacity-achieving pilot sequence design is used in the network. Importantly, when $N>0$, the SINR requirements may lie outside the per-cell user capacity region with $N=0$. Since the pilot sequences are designed under the assumption of $N=0$, the user capacity-achieving pilot sequence design can no longer guarantee to satisfy the SINR requirements for all Bobs in the network.

\begin{figure}[!t]
\centering
    \includegraphics[height=2.60in,width=0.95\columnwidth]{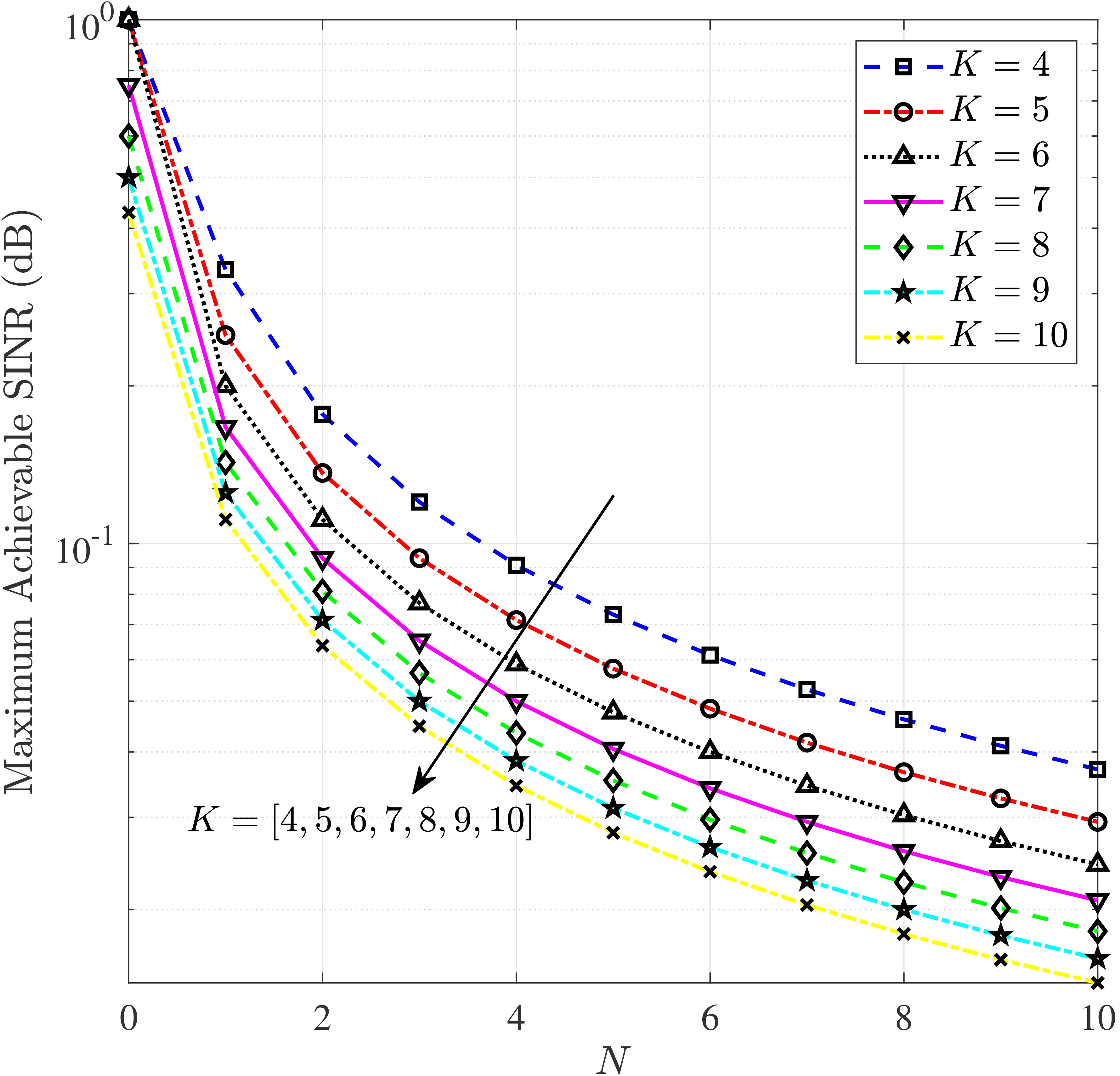}
    \caption{The maximum achievable SINR versus the number of Eves in the network for different number of Bobs in the network.}\label{maxSINRplot}
\end{figure}

We next evaluate the impact of increasing the number of Bobs $K$ and Eves $N$ on the maximum achievable SINR of the massive MIMO network. In this simulation, we assume that all Bobs have the same SINR requirements, i.e., $\gamma_{l1}=\cdots=\gamma_{lK}=\gamma$, where $\gamma$ is the maximum achievable SINR. Fig. \ref{maxSINRplot} depicts the maximum achievable SINR versus the number of Eves for different values of $K$. We note that increasing $N$ decreases the maximum achievable SINR in the network, which signifies the impact of Eves on the network. For example, when $K=4$ and $N$ increases from $N=0$ to $N=10$, the maximum achievable SINR reduces from $1.0$ to $0.03$, which amounts to $97\%$ reduction. We also note that increasing $K$ decreases the maximum achievable SINR in the network. For example, when $N=0$ and $K$ increases from $K=4$ to $K=10$, the maximum achievable SINR reduces from $1.0$ to $0.43$, which is equivalent to $57\%$ reduction in the maximum achievable SINR. Furthermore, the user capacity-achieving pilot sequence design is capable of designing pilot sequences to satisfy the SINR requirement of all Bobs in a cell or equivalently all Bobs in the network as long as the SINR requirements remain inside the per-cell user capacity region. As such, increasing $N$ adversely affects the network performance compared to increasing $K$.

\begin{figure}[!t]
\centering
    \includegraphics[height=2.60in,width=0.95\columnwidth]{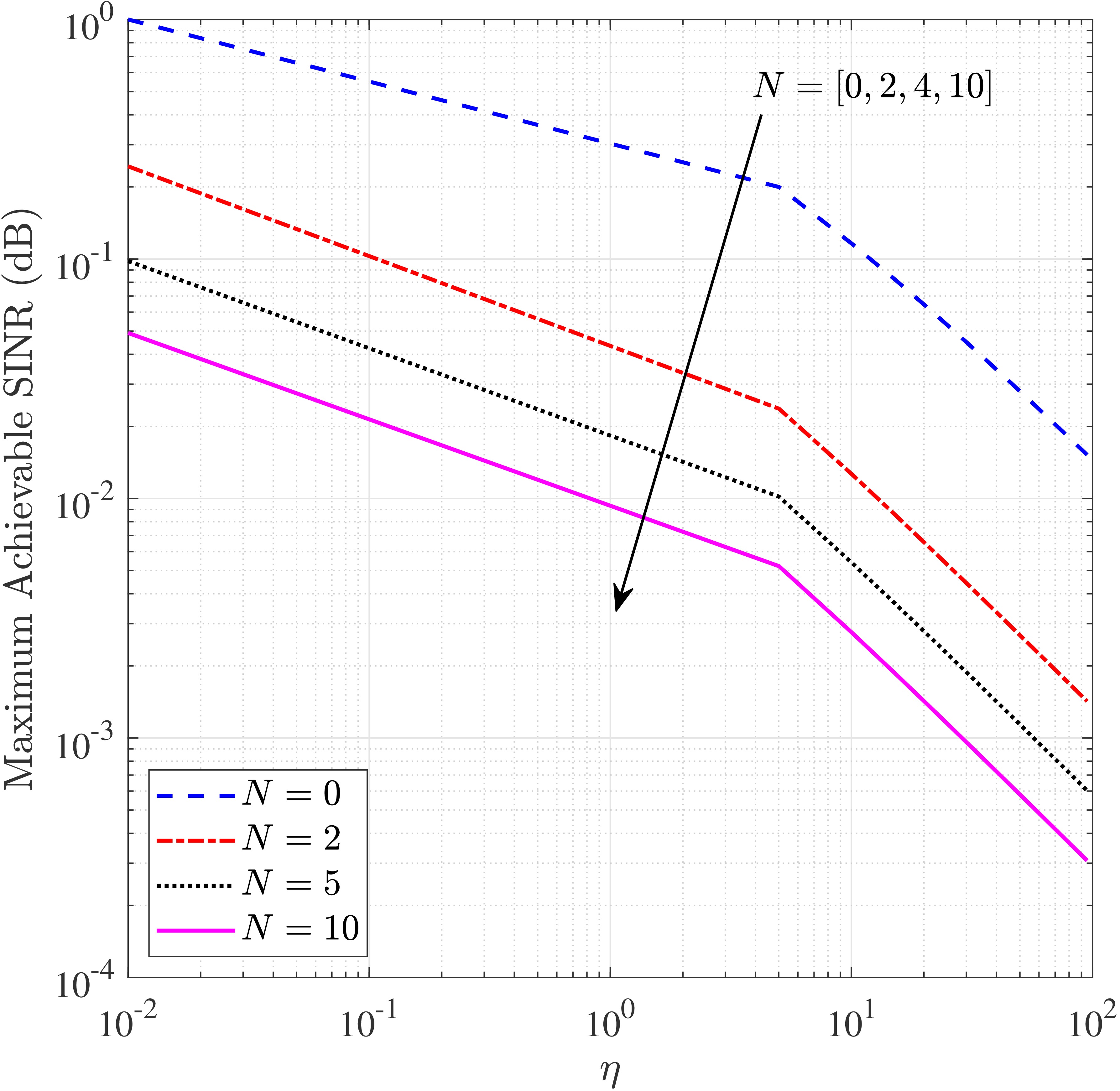}
    \caption{The maximum achievable SINR versus $\eta$ for different number of Eves in the network.}\label{maxSINRplot_percent}
\end{figure}

We next examine the impact of increasing the maximum achievable SINR requirements in the presence of $N$ Eves in the network. The SINR requirements in the network are given by $\bm{\gamma}_1=\left[\gamma, \gamma, \gamma, \eta\gamma\right]$ and $\bm{\gamma}_2=\left[\eta\gamma, \eta\gamma, \eta\gamma, \eta\gamma\right]$, where $\eta$ denotes a scaling factor applied to the SINR requirements for a few Bobs. Fig.~\ref{maxSINRplot_percent} shows the maximum achievable SINR versus $\eta$ for different values of $N$. We note that increasing the SINR requirements in the network, i.e., increasing $\eta$, decreases the maximum achievable SINR in the network. For example, when $N=2$ and $\eta$ increases from $0.01$ to $10$, the maximum achievable SINR decreases from $0.24$ to $0.001$, which represents a $99\%$ decrease. Moreoever, when $\eta=0.01$ and $N$ increases from $0$ to $10$, the maximum achievable SINR decreases from $1.0$ to $0.05$, which represents a $95\%$ decrease. We highlight that for $\eta=0.01$, the increase in $N$ from $0$ to $2$ means that the Bobs with SINR requirements above $0.04$ can no longer be satisfied using the user capacity-achieving pilot sequence design.

\begin{figure}[!t]
\centering
    \includegraphics[height=2.65in,width=0.95\columnwidth]{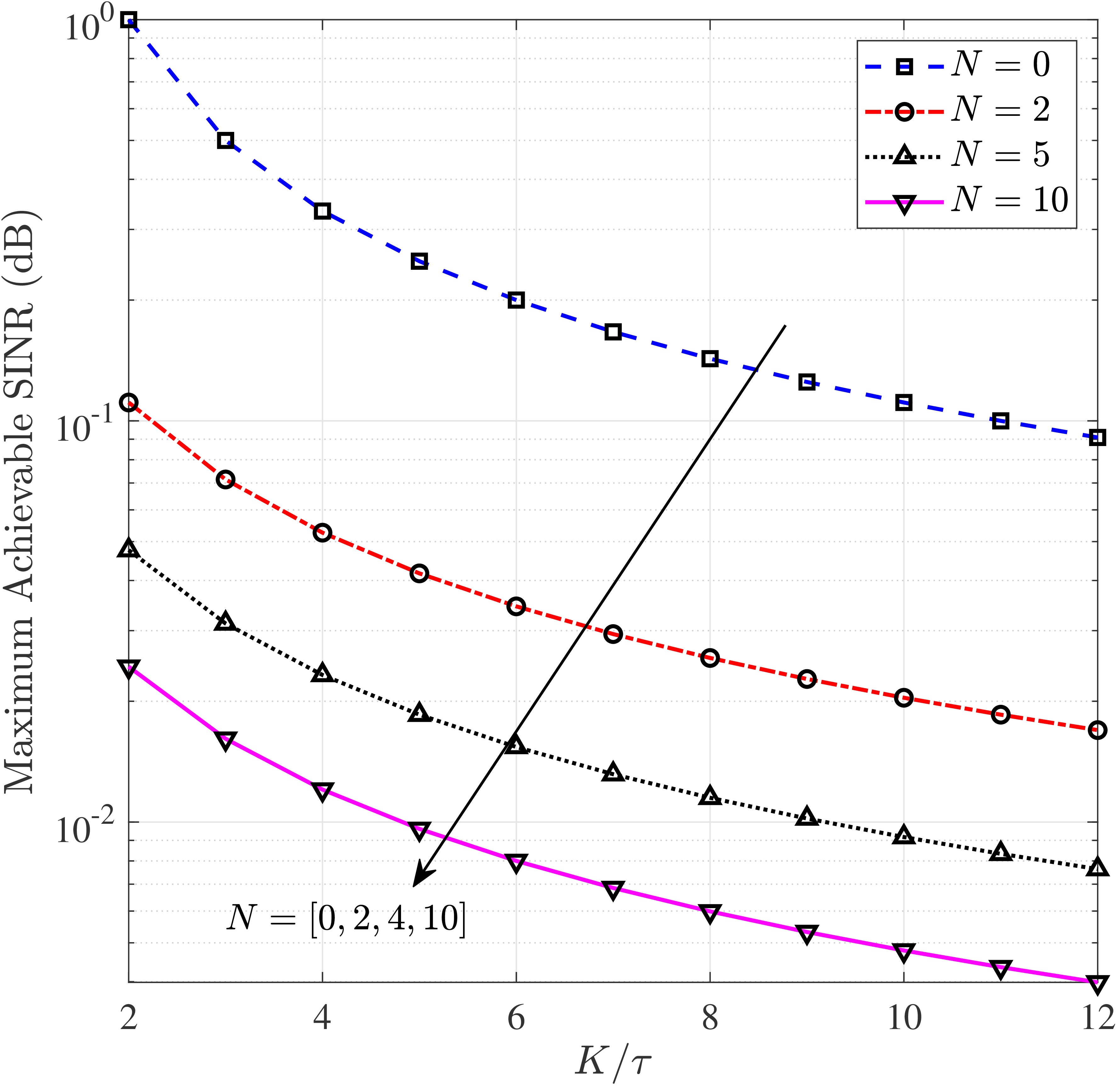}
    \caption{The maximum achievable SINR versus the pilot loading factor for different number of Eves in the network.}\label{pilot_loading}
\end{figure}

We now evaluate the maximum achievable SINR against the pilot loading factor \footnote{In literature, pilot loading factor is used for the orthogonal pilot sequences. A pilot loading factor of $K/\tau = 2$ represents that two Bobs in a cell are assigned the same pilot sequence.}, which is defined as the ratio between the number of Bobs and the length of the pilot sequence, i.e., $K/\tau$. We highlight that pilot loading factor is a fair parameter for performance evaluation since a higher pilot loading factor represents a higher correlation between various pilot sequences. In other words, the pilot sequences designed by the user capacity-achieving pilot sequence design are less orthogonal. Fig.~\ref{pilot_loading} depicts the maximum achievable SINR versus the pilot loading factor for different values of $N$. We note that increasing the pilot loading factor decreases the maximum achievable SINR in the network. For example, when $N=2$ and $K/\tau$ increases from $2$ to $8$, the maximum achievable SINR reduces from $0.1$ to $0.03$. However, the impact of increasing $N$ is much larger than the impact of increasing $K/\tau$. For example, when $N$ increases from $0$ to $2$, the decrease in maximum achievable SINR is comparable with that when $N=0$ and $K/\tau$ increases from $2$ to $8$. This performance comparison highlights the degradation in the achievable SINR caused by Eves.

\subsection{Impact of PC Attack on Achievable SINRs}

In this subsection, we examine the impact of the PC attack on the achievable SINR for Bobs in the $l$-th cell of the network, when $\textrm{Alice}_l$ is equipped with a large but finite number of antennas. In this subsection, we assume that the SINR requirements for all Bobs in the network are given as $\bm{\gamma}_1 = [0.91, 0.74, 0.64, 0.23]$ and $\bm{\gamma}_2 = [0.94, 0.82, 0.45, 0.20]$. Accordingly, we use the user capacity-achieving pilot sequence design to obtain pilot sequences for all Bobs in the network. We highlight that the SINR requirements lie inside the user capacity region given by \eqref{BW_all_eve}, when $N=0$.

\begin{figure}[!t]
\centering
    \includegraphics[height=2.65in,width=0.95\columnwidth]{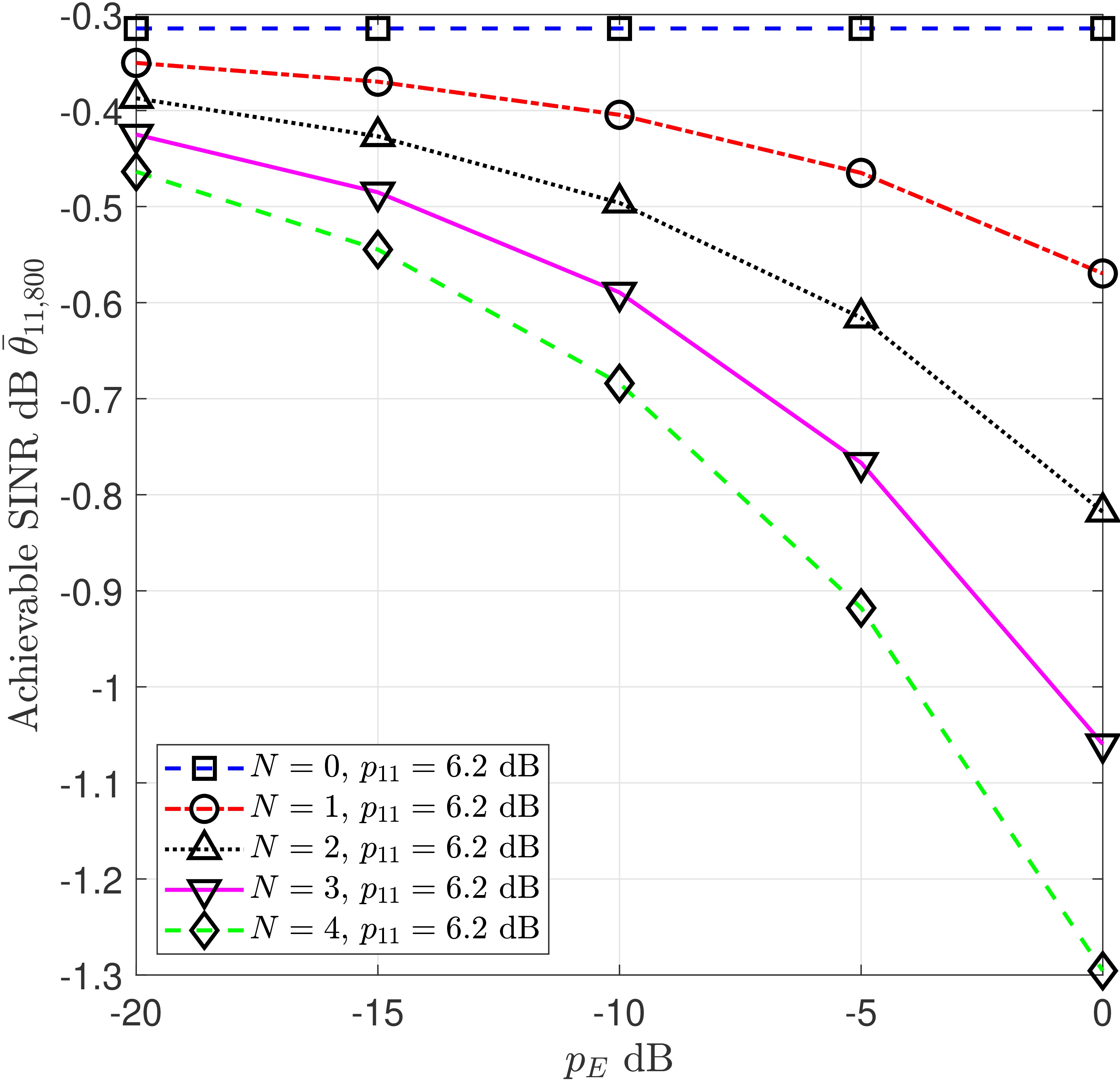}
    \caption{The achievable SINR $\bar{\theta}_{11,800}$ versus the transmit power of $\textrm{Eve}_n$ for different number of Eves in the network.}\label{m_power}
\end{figure}

We first examine the impact of increasing the transmit power of $\textrm{Eve}_{in}$, $\bar{p}_{in}$, on the achievable SINR. We highlight that $\bar{p}_{in}$ cannot exceed a certain threshold, which depends on network conditions, the power resource available at $\textrm{Eve}_n$, and the detection capabilities of Alice. Fig.~\ref{m_power} depicts the achievable SINR of $\textrm{Bob}_{11}$, $\bar{\theta}_{11,800}$, versus $\bar{p}_{in}$ for different values of $N$. In this figure, we assume that $\textrm{Alice}_l$ is equipped with $800$ antennas and $\bar{p}_{in}$ must satisfy the constraint given by \eqref{pcpower}. We find that increasing $\bar{p}_{in}$ reduces $\bar{\theta}_{11,800}$. Moreover, we find that the active attack strategy is able to reduce $\bar{\theta}_{11,800}$ by $23.8\%$ when $\bar{p}_{in}=0~\textrm{dB}$, which implies that $\bar{p}_{in}$ is approximately $49\%$ of the maximum transmit power of Bobs in the network. Importantly, we observe the reduction in $\bar{\theta}_{11,800}$ even when the transmit power is very small, e.g., $\bar{p}_{in}=-20~\textrm{dB}$, when $N=1$. As such, the user capacity-achieving pilot sequence design is adversely affected by the active attack even when the attacker has limited resources. Throughout the remainder of this subsection, we assume that $\bar{p}_{in}=0~\textrm{dB}$.

\begin{figure}[!t]
\centering
    \includegraphics[height=2.65in,width=0.95\columnwidth]{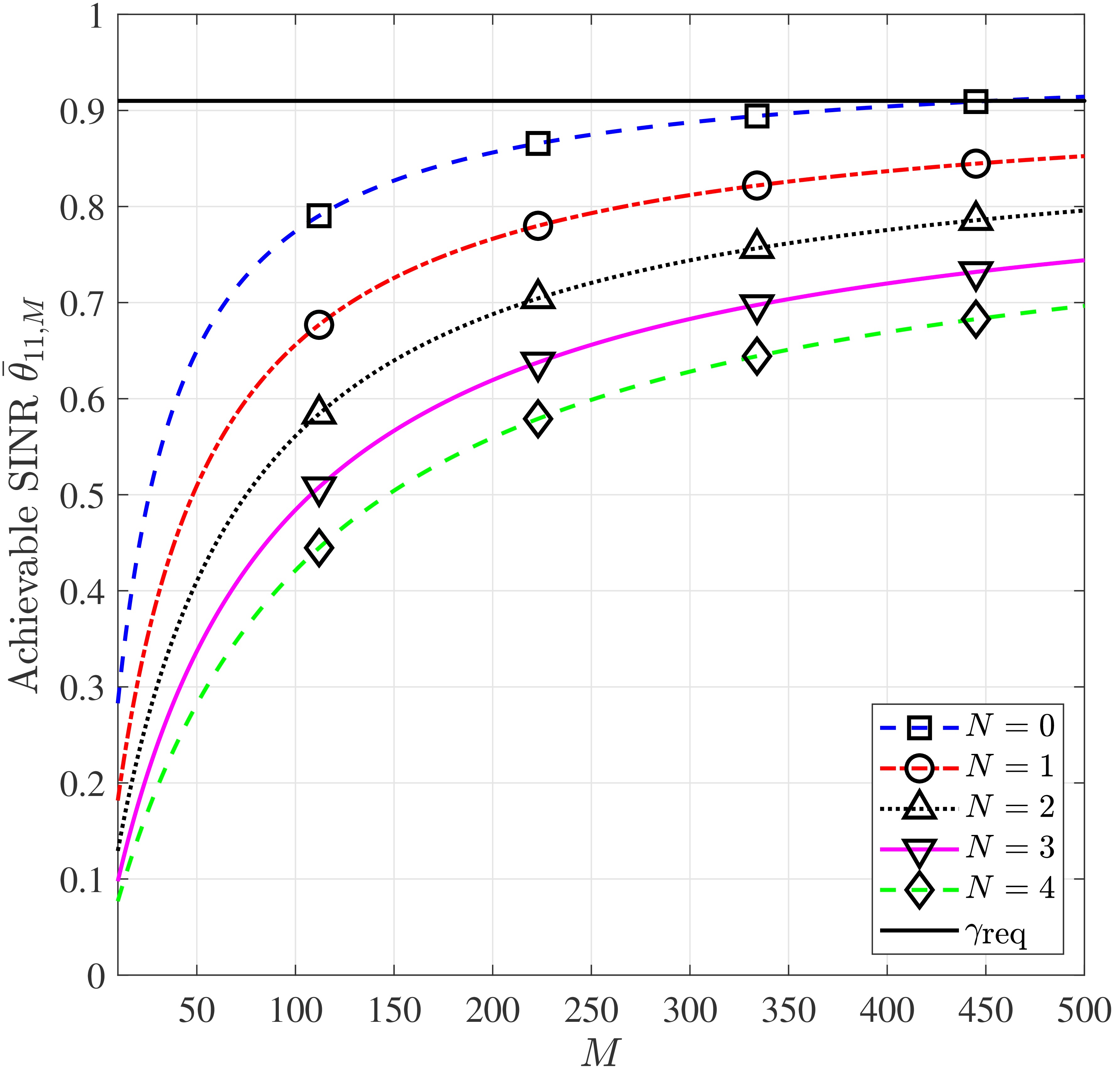}
    \caption{The achievable SINR $\bar{\theta}_{11,M}$ versus the number of antennas at the BS for different number of Eves in the network.}\label{SINR_11}
\end{figure}

We next examine the impact of the active attack strategy employed by Eves on the achievable SINR for $\textrm{Bob}_{11}$. Fig.~\ref{SINR_11} depicts $\bar{\theta}_{11,M}$ versus the number of antennas at $\textrm{Alice}_l$ for different values of $N$. This figure highlights the severity of the active attack strategy on the network performance. For example, $\textrm{Bob}_{11}$ can no longer satisfy its SINR requirements even when there is one Eve in the network. Importantly, when $M=500$ and $N=1$, the achievable SINR $\bar{\theta}_{11,M}$ reduces from $0.91$ to $0.85$, which is equivalent to $10.9\%$ reduction in the achievable SINR. Consequently, $\bar{\theta}_{11,M}$ can no longer be satisfied even with an unlimited number of antennas at the BS. Furthermore, we find that $\bar{\theta}_{11,M}$ reduces by $23.08\%$ when $N=3$ as compared to $\bar{\theta}_{11,M}$ when $N=0$. As such, the active attack strategy effectively reduces the achievable SINR of Bobs in the network where the pilot sequences are designed by using the user capacity-achieving pilot sequence design.

\begin{figure}[!t]
\centering
    \includegraphics[height=2.65in,width=0.95\columnwidth]{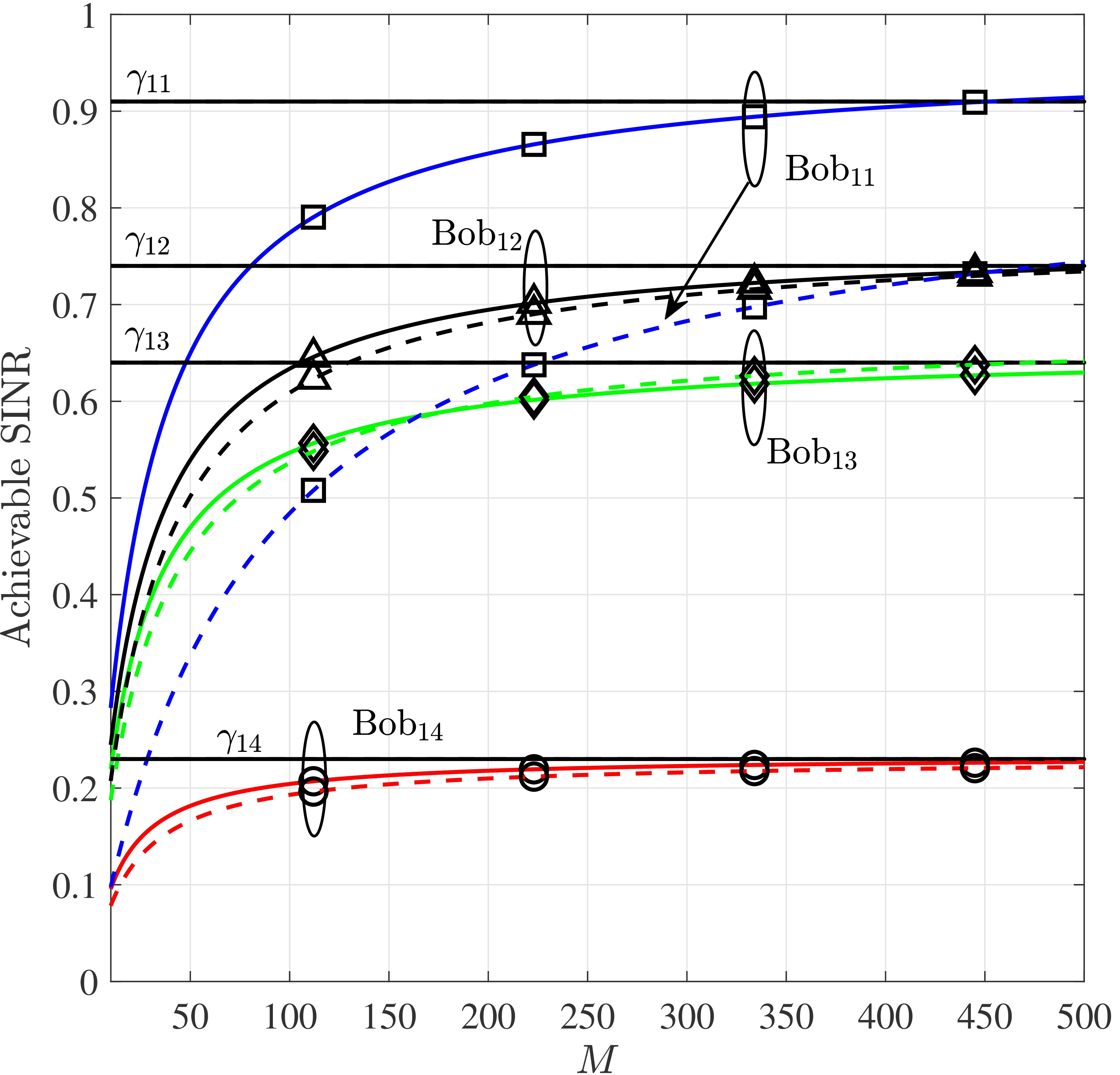}
    \caption{The achievable SINR $\bar{\theta}_{11,M}$, $\bar{\theta}_{12,M}$, $\bar{\theta}_{13,M}$, and $\bar{\theta}_{14,M}$ versus the number of antennas for $N=0$ and $N=3$.}
\label{m_antennaSINR18}
\end{figure}

Finally, we examine the impact of active attack on achievable SINR of Bobs in the network. We highlight that the pilot sequences used in the network are correlated. As such, increased PC that is caused by the active attack has the potential to affect the achievable SINR of more than one Bob in the network. Fig.~\ref{m_antennaSINR18} depicts the achievable SINR for $\textrm{Bob}_{11}$, $\textrm{Bob}_{12}$, $\textrm{Bob}_{13}$, and $\textrm{Bob}_{14}$. In this simulation, the solid curves are obtained from \eqref{SINR_eve} when $N=0$ and dashed curves are obtained from \eqref{SINR_eve} when $N=3$. We find that the active attack strategy reduces the achievable SINR for all Bobs in the network except for $\textrm{Bob}_{12}$. Notably, the achievable SINR for $\textrm{Bob}_{11}$ is significantly reduced because its pilot sequence is known to Eves. We note that the active attack strategy not only impacts the SINR requirements for the Bobs whose pilot sequence is known to Eves but impacts the achievable SINR of other Bobs. We highlight that the use of correlated pilots is the main cause of this impact. Thus, correlated pilots design such as the user capacity-achieving pilot sequence design may be more suitable for a network without Eves. Furthermore, we note that the difference between the achievable SINR without and with Eves in the network decreases as $M$ increases. For example, when $M=100$, the achievable SINR for $\textrm{Bob}_{11}$ reduces $36.81\%$ due to active attack. However, when $M=800$ the achievable SINR for $\textrm{Bob}_1$ reduces $18.69\%$ due to active attack. This behaviour is due to the large array gain provided by massive MIMO, which indicates the potential benefit of using massive MIMO in a network vulnerable to active attacks.

\section{Conclusion}\label{conclusion}

In this paper, we proposed an active attacking strategy on the user capacity-achieving pilot sequence design in a massive MIMO network, where correlated pilots are assigned to all the users in the network. To this end, we first identified the potential vulnerabilities in the user capacity-achieving pilot sequence design which produces correlated pilots for all the users in the network. Based on the known structure of the correlated pilot set, we then designed an active attacking strategy for increasing PC in the massive MIMO network. We demonstrated that the SINR requirements for all the users in the network are no longer guaranteed in the presence of active attacker(s) using our proposed strategy. The degradation in network performance is due to the fact that the user capacity region is significantly reduced by active attackers. Specifically, the SINR requirements for the worst-affected users by active attackers may not be satisfied even with an infinite number of antennas at the BS. We concluded that the use of correlated pilots make the massive MIMO network prone to active attack due to increased PC.

\appendices

\section{Proof of Proposition~\ref{prop_K_tot}}\label{appa}
In this appendix, we ascertain the impact of Eves' PC attack on the user capacity of the network. Assuming that $\textrm{Alice}_l$ is equipped with a large number of antennas, i.e., $M\rightarrow\infty$, and $p_{ij}\beta_{lki} \leq 1$ we obtain from \eqref{SINR_eve} that
\begin{align}\label{SINR_eve_infty}
\bar{\theta}_{lk,\infty}&=\frac{P_{lk}}{\bar{\alpha}_{lk}\sum_{(i,j)\neq (l,k)}\frac{\left|\rho_{lkij}\right|^2 P_{ij}}{\bar{\alpha}_{ij}}}\notag\\
&=\frac{P_{lk}}{\bar{\alpha}_{lk}\sum_{i,j}\frac{\left|\rho_{lkij}\right|^2 P_{ij}}{\bar{\alpha}_{ij}}-P_{lk}}.
\end{align}
We highlight that the downlink achievable SINR with an infinite number of antennas $\bar{\theta}_{lk,\infty}$,  is always greater than or equal to the downlink achievable SINR with a finite number of antennas $M$, $\theta_{lk,M}$. From \eqref{alpha_corr_eve} and \eqref{alpha_corr}, we obtain
\begin{align}\label{SINR_eve_infty2}
\bar{\alpha}_{lk}=\alpha_{lk}+\hat{\alpha}_{11lk}+\cdots+\hat{\alpha}_{inlk}+\cdots+\hat{\alpha}_{LNlk},
\end{align}
where $\hat{\alpha}_{inlk}=\bar{p}_{in}\bar{\beta}_{inl}\rho_{inlk}^{2}$ and $\bar{p}_{n}>0$. Using \eqref{SINR_eve_infty} and \eqref{SINR_eve_infty2}, we obtain
\begin{align}\label{SINR_eve_infty3}
\bar{\theta}_{lk,\infty}&=\frac{P_{lk}}{\left(\alpha_{lk}+\hat{\alpha}_{11lk}+\cdots+\hat{\alpha}_{LNlk}\right)\sum_{i,j} \frac{\left|\rho_{lkij}\right|^2 P_{ij}}{\bar{\alpha}_{ij}}-P_{lk}},\notag\\
&=\frac{P_{lk}}{\left(\alpha_{lk}+\cdots+\hat{\alpha}_{LNlk}\right)\textrm{tr}\left(\mathbf{s}_{lk}^{T}\mathbf{S}\mathbf{P} \mathbf{\bar{A}} \mathbf{S}^{T}\mathbf{s}_{lk}\right)-P_{lk}},
\end{align}
where we define the matrices $\mathbf{S}=\left[\mathbf{S}_{1}, \cdots, \mathbf{S}_{l},\cdots, \mathbf{S}_{L}\right]$, $\mathbf{P}=\text{diag}\left[\mathbf{P}_{1},\cdots,\mathbf{P}_l,\cdots, \mathbf{P}_{L}\right]$, and $\mathbf{\bar{A}}=\text{diag}\left[\mathbf{\bar{A}}_1,\cdots, \mathbf{\bar{A}}_l, \cdots, \mathbf{\bar{A}}_L\right]$. Furthermore, we define $\mathbf{S}_l=\left[\mathbf{s}_{l1}, \mathbf{s}_{l2},\cdots, \mathbf{s}_{lK}\right]$, $\mathbf{P}_l=\textrm{diag}\left[{p}_{l1}, {p}_{l2},\cdots, {p}_{lK}\right]$, and $\mathbf{\bar{A}}_l=\textrm{diag}\left[\frac{1}{\bar{\alpha}_{l1}}, \frac{1}{\bar{\alpha}_{l2}},\cdots, \frac{1}{\bar{\alpha}_{lK}}\right]$ respectively. From \eqref{SINR_eve_infty3}, we obtain
\begin{align}\label{SINR_eve_infty4}
\sum_{l,k}\frac{1+\bar{\theta}_{lk,\infty}}{\bar{\theta}_{lk,\infty}}=
&\sum_{l,k} \frac{\left(\alpha_{lk}+\cdots+\hat{\alpha}_{LNlk}\right)
\textrm{tr}\left(\mathbf{s}_{lk}^{T}\mathbf{S}\mathbf{P}\mathbf{\bar{A}} \mathbf{S}^{T}\mathbf{s}_{lk}\right)}{P_{lk}}\notag\\
=&\underbrace{\textrm{tr}\left(\mathbf{P}^{-1}\mathbf{A}^{-1}\mathbf{S}^T \mathbf{S}\mathbf{P}\mathbf{\bar{A}}\mathbf{S}^{T}\mathbf{S}\right)}_{\upsilon_{1}}+\cdots\notag\\ &+\underbrace{\textrm{tr}\left(\mathbf{P}^{-1}\bar{\mathbf{A}}_{LN}^{-1}\mathbf{S}^T \mathbf{S}\mathbf{P}\mathbf{\bar{A}}\mathbf{S}^{T}\mathbf{S}\right)}_{\upsilon_{2}},
\end{align}
where we use the matrix definitions $\mathbf{{A}}= \textrm{diag}[\mathbf{{A}}_1,\cdots, \mathbf{{A}}_l,\cdots, \mathbf{{A}}_L]$,  $\mathbf{{A}}_l = \text{diag}\left[\frac{1}{{\alpha}_{l1}},\frac{1}{{\alpha}_{l2}},\cdots,\frac{1}{{\alpha}_{lK}}\right]$ , and $\mathbf{\bar{A}}_{in}$ is a block diagonal matrix with diagonal entries given as $\mathbf{\bar{A}}_{in}= \textrm{diag}[\mathbf{\bar{A}}_{in1},\cdots, \mathbf{\bar{A}}_{inl},\cdots, \mathbf{\bar{A}}_{inL}]$ and $\bar{\mathbf{A}}_{inl}= \text{diag}\left[\frac{1}{\hat{\alpha}_{inl1}},\frac{1}{\hat{\alpha}_{inl2}},\cdots,\frac{1}{\hat{\alpha}_{inlK}}\right]$. We note that $\upsilon_{2}$ in \eqref{SINR_eve_infty4} is due to PC attack by $\textrm{Eve}_{LN}$. We next simplify $\upsilon_{1}$ in \eqref{SINR_eve_infty4} as
\begin{align}\label{interme-appendix7}
\upsilon_{1}
&=K_{\textrm{tot}}+\sum_{p,q \atop p > r}\sum_{r,s \atop q > s}\left(\frac{\alpha_{rs}P_{pq}}{\bar{\alpha}_{pq}P_{rs}} +\frac{\bar{\alpha}_{pq}P_{rs}}{\alpha_{rs}P_{pq}}\right)\left|\rho_{pqrs}\right|^2\notag\\
&\geq K_{\textrm{tot}} + 2\sum_{p,q \atop p > r}\sum_{r,s \atop q > s}\left|\rho_{pqrs}\right|^2\notag\\
&\geq \textrm{tr}\left(\mathbf{R}_{S}\mathbf{R}_{S}\right),
\end{align}
where $\mathbf{R}_{S}=\mathbf{S}^T\mathbf{S}$. Using the eigen decomposition of $\mathbf{R}_{S}$, we rewrite \eqref{interme-appendix7} as
\begin{align}\label{interme-appendix7a}
\textrm{tr}\left(\mathbf{R}_{S}\mathbf{R}_{S}\right)= \sum_{i=1}^{K_{\textrm{tot}}}\lambda_{i}^{2}=\frac{\left(\sum_{i=1}^{\tau}\lambda_{i}\right)^{2}}{\tau} =\frac{K_{\textrm{tot}}^{2}}{\tau},
\end{align}
where $\lambda_{i}$ is the $i$-th eigen value of $\mathbf{R}_{S}$. From \eqref{interme-appendix7} and \eqref{interme-appendix8}, we obtain
\begin{align}\label{interme-appendix8}
\upsilon_{1}\geq\frac{K_{\textrm{tot}}^2}{\tau}.
\end{align}
Following the same procedure, we obtain
\begin{align}\label{interme-appendix9}
\upsilon_{2}\geq\frac{K_{\textrm{tot}}^2}{\tau}.
\end{align}
Finally, substituting \eqref{interme-appendix8} and \eqref{interme-appendix9} in \eqref{SINR_eve_infty4}, we obtain the bound on the user capacity of the massive MIMO network in the presence of $N$ Eves as
\begin{align}\label{interme-appendix10}
\sum_{l,k}\frac{1+\bar{\theta}_{lk,\infty}}{\bar{\theta}_{lk,\infty}}
\geq\frac{\left(LN+1\right)K_{\textrm{tot}}^{2}}{\tau}.
\end{align}
Noting the $\bar{\theta}_{lk,\infty}\geq\gamma_{lk}$, we then rewrite \eqref{interme-appendix10} as
\begin{align}\label{interme-appendix101}
\sum_{l=1}^L\sum_{k=1}^K\frac{1+\gamma_{lk}}{\gamma_{lk}}\geq\frac{\left(LN+1\right)K_{\textrm{tot}}^{2}}{\tau},
\end{align}
which is then rearranged to obtain \eqref{K_to_eve}.

\begin{IEEEbiography}[{\includegraphics[width=1in,height=1.25in,clip,keepaspectratio]{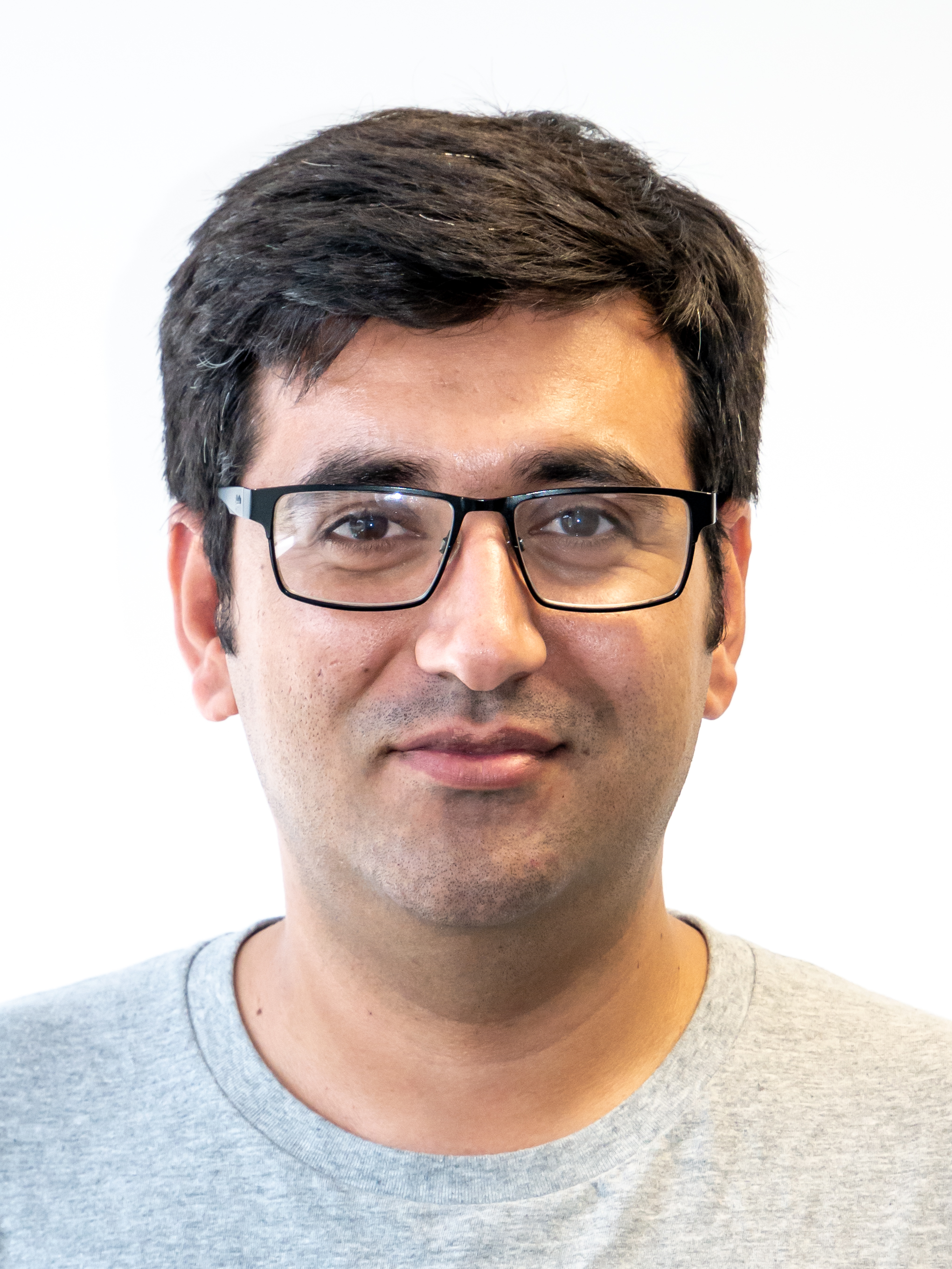}}]%
{Noman Akbar} (S’15--M'20) received the B.E. degree in electrical (telecommunication) engineering from the National University of Sciences and Technology (NUST), Rawalpindi, Pakistan, in 2012, the M.S. degree in computer engineering from Kyung Hee University, Suwon, South Korea, in 2015, and  the Ph.D. degree in electrical engineering from the Australian National University, Canberra, Australia, in 2018.  From 2012 to 2013, he was a Researcher with the NUST School of Electrical Engineering and Computer Science. From 2013 to 2015, he was with the Haptics Laboratory, Kyung Hee University, Suwon, South Korea, as a Research Assistant. He is currently a Postdoctoral Research Fellow with the Research School of Electrical, Energy and Materials Engineering, Australian National University, Canberra, Australia. He received the Best Paper Award from IEEE GlobeCOM in 2016. His research interests include signal processing for audio and wireless communications.
\end{IEEEbiography}
\vspace{-0.4in}
\begin{IEEEbiography}[{\includegraphics[width=1in,height=1.25in,clip,keepaspectratio]{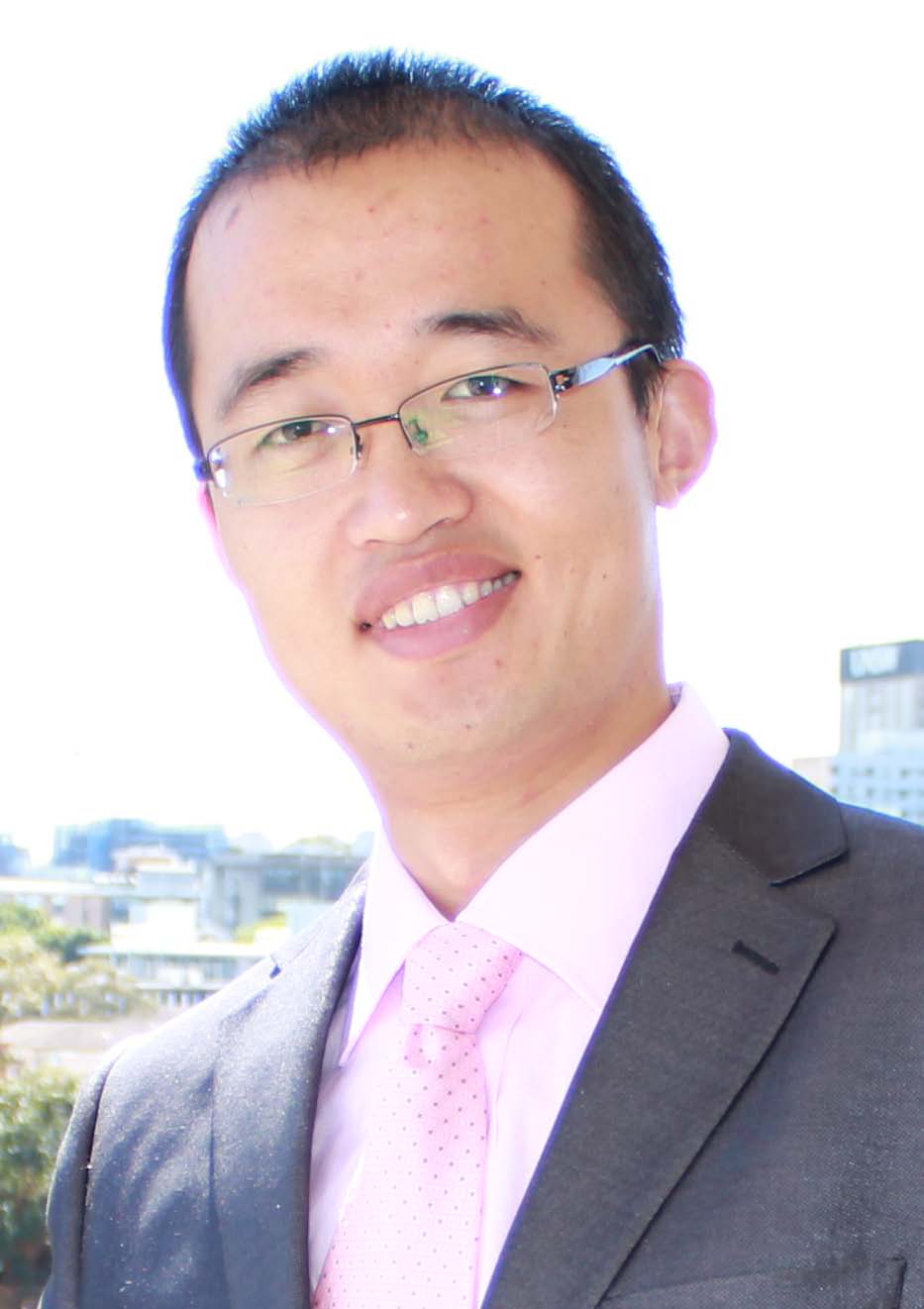}}]%
{Shihao Yan} (S’11--M’15) received the B.S. degree in communication engineering and the M.S. degree in communication and information systems from Shandong University, Jinan, China, in 2009 and 2012, respectively, and the Ph.D. degree in electrical engineering from the University of New South Wales, Sydney, NSW, Australia, in 2015. From 2015 to 2017, he was a Postdoctoral Research Fellow in the Research School of Engineering, Australian National University, Canberra, ACT, Australia. He is currently a University Research Fellow with the School of Engineering, Macquarie University, Sydney, NSW, Australia. His current research interests are in the areas of wireless communications and statistical signal processing, including physical layer security, covert communications, and location spoofing detection.
\end{IEEEbiography}
\vspace{-0.4in}
\begin{IEEEbiography}[{\includegraphics[width=1in,height=1.25in,clip,keepaspectratio]{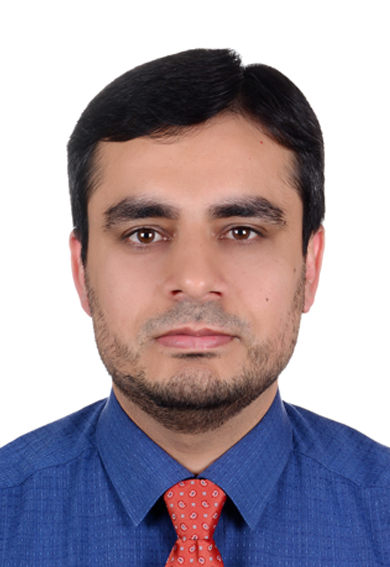}}]%
{Asad Masood Khattak} is an Associate Professor at the College of Technological Innovation, Zayed University in Abu Dhabi, UAE that he joined in August 2014. He received his M.S. in Information Technology from National University of Sciences and Technology (NUST), Islamabad, Pakistan in 2008 and received his Ph.D. degree in Computer Engineering from Kyung Hee University, South Korea in 2012. He worked as Post-Doctoral Fellow at Department of Computer Engineering, Kyung Hee University, South Korea and later joined the same college as Assistant Professor. He is currently leading three research projects, collaborating in four research projects, and has successfully completed five research projects in the fields of Data Curation, Context-aware Computing, IoT, and Secure Computing. He is an IEEE member and has authored/coauthored more than 100 journal and conference articles in highly reputed venues. He is serving as reviewer, program committee member, and guest editor of many conferences and journals. He has delivered keynote speeches, invited talks, guest lectures and has delivered short courses in many universities. He and his team have secured several national and international awards in different competitions.
\end{IEEEbiography}
\vspace{2.5in}
\begin{IEEEbiography}[{\includegraphics[width=1in,height=1.25in,clip,keepaspectratio]{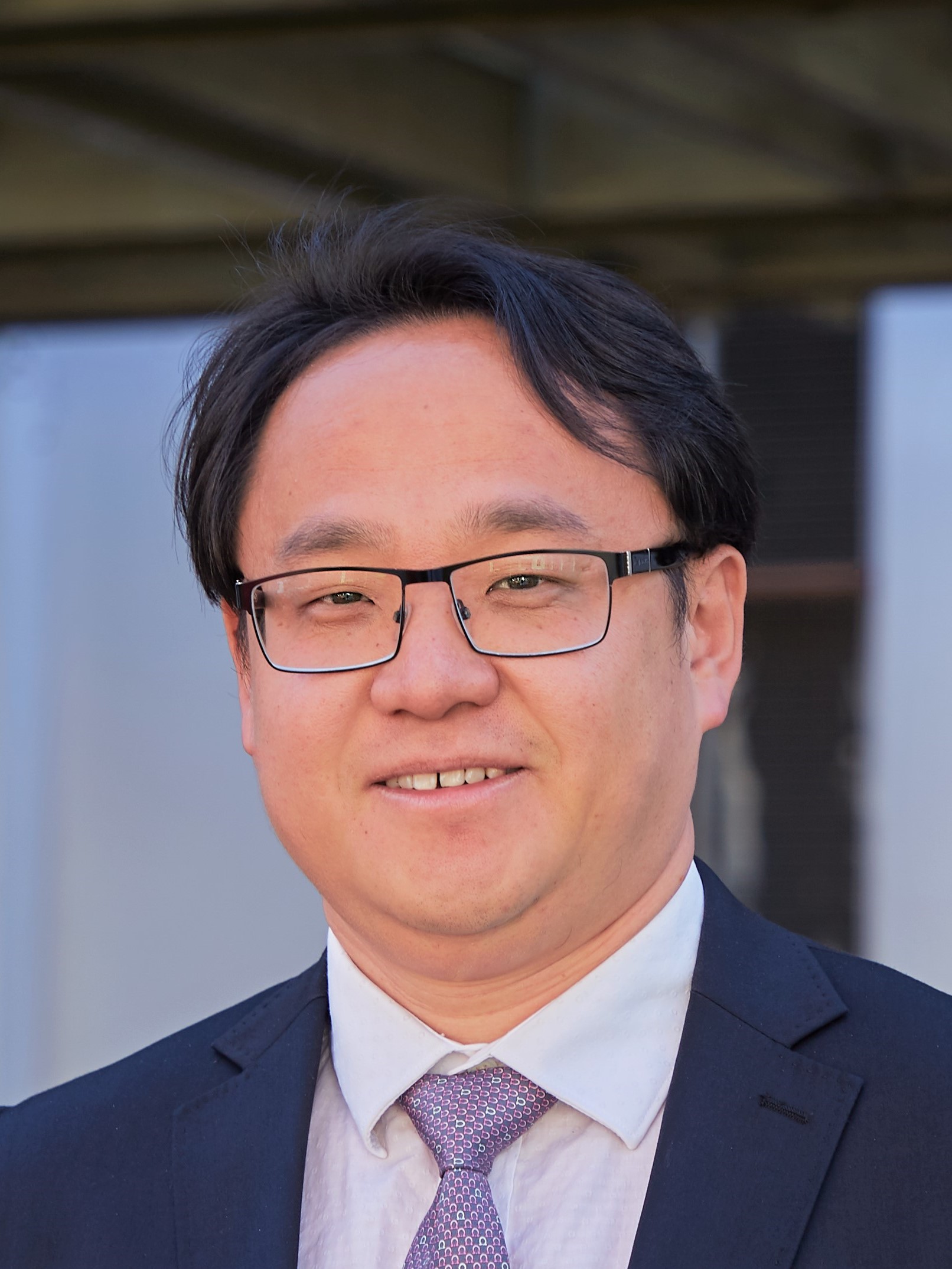}}]
{Nan Yang} (S'09--M'11--SM'18) received the B.S. degree in electronics from China Agricultural University in 2005, and the M.S. and Ph.D. degrees in electronic engineering from the Beijing Institute of Technology in 2007 and 2011, respectively. He has been with the Research School of Electrical, Energy and Materials Engineering at the Australian National University since July 2014, where he currently works as a Senior Lecturer. Prior to this, he was a Postdoctoral Research Fellow at the University of New South Wales (2012--2014) and the Commonwealth Scientific and Industrial Research Organization (2010--2012). He received the IEEE ComSoc Asia-Pacific Outstanding Young Researcher Award in 2014 and the Best Paper Awards from the IEEE GlobeCOM 2016 and the IEEE VTC 2013-Spring. He also received the Top Editor Award from the \textsc{Transactions on Emerging Telecommunications Technologies}, the Exemplary Reviewer Awards from the \textsc{IEEE Transactions on Communications}, \textsc{IEEE Wireless Communications Letters}, and \textsc{IEEE Communications Letters}, and the Top Reviewer Award from the \textsc{IEEE Transactions on Vehicular Technology} from 2012 to 2018. He is currently serving in the Editorial Board of the \textsc{IEEE Transactions on Wireless Communications}, \textsc{ IEEE Transactions on Molecular, Biological, and Multi-Scale Communications}, \textsc{IEEE Transactions on Vehicular Technology}, and \textsc{Transactions on Emerging Telecommunications Technologies}. His general research interests include ultra-reliable low latency communications, millimeter wave and terahertz communications, massive multi-antenna systems, cyber-physical security, and molecular communications.
\end{IEEEbiography}

\end{document}